\documentclass[prd,twocolumn,amsmath,amssymb,floatfix,superscriptaddress]{revtex4-1}

\usepackage{graphicx}
\usepackage{amssymb}
\usepackage{hyperref}
\usepackage{color}

\definecolor{darkgreen}{cmyk}{0.85,0.2,1.00,0.2} 
\definecolor{purple}{cmyk}{0.5,1.0,0,0}

\newcommand{\fstr}{f_{\rm str}}

\def\barray{\begin{array}}
\def\earray{\end{array}}
\def\be{\begin{equation}} 
\def\ee{\end{equation}}
\def\ben{\begin{equation} \nonumber}
\def\een{\end{equation}}
\def\ban{\begin{eqnarray*}}
\def\ean{\end{eqnarray*}}
\def\ba{\begin{eqnarray}}
\def\ea{\end{eqnarray}}

\def\({\left(}
\def\){\right)}

\begin{document}

\title{Cosmic String constraints from WMAP and the South Pole Telescope}
\author{Cora Dvorkin}

\affiliation{Kavli Institute for Cosmological Physics, University of Chicago, Chicago IL 60637}
\affiliation{Department of Physics, University of Chicago, Chicago IL 60637}
\affiliation{School of Natural Sciences, Institute for Advanced Study, Princeton, NJ 08540}

\author{Mark Wyman}

\affiliation{Kavli Institute for Cosmological Physics, University of Chicago, Chicago IL 60637}
\affiliation{Department of Astronomy \& Astrophysics, University of Chicago, Chicago IL 60637}

\author{Wayne Hu}

\affiliation{Kavli Institute for Cosmological Physics, University of Chicago, Chicago IL 60637}
\affiliation{Department of Astronomy \& Astrophysics, University of Chicago, Chicago IL 60637}


\begin{abstract}
The predictions of the inflationary $\Lambda$CDM paradigm  match today's high-precision 
measurements of the cosmic microwave background anisotropy extremely well. The same data
put tight limits on other sources of anisotropy.
Cosmic strings are a particularly interesting alternate source to constrain. Strings are topological defects, remnants of inflationary-era
physics that persist after the big bang. They are formed in a variety of models of inflation, including 
string theory models such as brane inflation. We assume
a ``Nambu-Goto" model for strings, approximated by a collection of unconnected segments with zero width, and show that measurements of temperature
anisotropy by the South Pole Telescope break a parameter degeneracy in the WMAP data, permitting us to place
 a strong upper limit on the possible string contribution to the CMB anisotropy: the  power sourced by zero-width strings must be $<1.75\%$ (95\% CL) of the total or the string tension $G\mu <1.7 \times10^{-7}$.
These limits imply that the best hope for detecting strings in the CMB will come from B-mode polarization measurements at arcminute scales rather than the degree scale measurements pursued for gravitational wave detection.
\end{abstract}
\maketitle

\section{Introduction}
An array of observations have now confirmed the predictions of the inflationary paradigm in its simplest form.
This leads us to ask what new physics could still be found by even more precise future experiments. Examples
of new sources for observables that could give great insight into fundamental physics include gravity waves \cite{cmbpol}, 
primordial non-Gaussianity \cite{nongauss}, and cosmic (super)strings \cite{Polchinski:2004ia, Myers:1900zz, Copeland:2011dx}. Cosmic strings are linear
topological defects formed whenever inflation ends with a phase transition that breaks a (gauged) U(1) symmetry.
Such a phase transition is expected in inflationary models motivated by supersymmetry ({\it e.g.}\ \cite{Cui:2007js})
and in string theory models like brane inflation \cite{Sarangi:2002yt}. 

Originally proposed as the origin of cosmological structure,  string contributions
to the CMB anisotropy have long been constrained to be less than 10\% of the total \cite{Albrecht:1997nt, Pogosian:2003mz, Pogosian:2004ny, Wyman:2005tu}.
Even at a reduced amplitude, strings remain interesting to study.
They could, for instance, be a source of observable B-mode polarization \cite{Seljak:1997ii, Battye:1998js, Benabed:1999wn, Bevis:2007qz, Pogosian:2007gi, Avgoustidis:2011ax},
gravitational waves \cite{Damour:2000wa, Olmez:2010bi, Leblond:2009fq}, $21$ cm radiation \cite{Khatri:2008zw,Berndsen:2010xc,Brandenberger:2010hn,Hernandez:2011ym}
or gravitational lensing \cite{Mack:2007ae, Chernoff:2007pd}.
A definitive detection of strings would greatly narrow down the inflationary parameter space and could be the only direct observational
window into the very high energy physics that generates strings.

The South Pole Telescope (SPT) has recently released a high-$\ell$ power spectrum
 of the CMB temperature anisotropy of unprecedented accuracy  \cite{Keisler:2011aw}.
 The time is therefore ripe to implement the test proposed in \cite{Pogosian:2008am}. In that paper, the authors argued
that strings should be either detected or strongly constrained by measurements of high-resolution CMB experiments like
the Atacama Cosmology Telescope (ACT) \cite{Hlozek:2011pc} or the SPT. This is because string contributions to the temperature anisotropy in the CMB fall off
 much more gradually at small angular scales than those sourced by inflation. 
 
 This weaker damping is a result of the very different way in which strings source
CMB anisotropy as compared with inflation. Inflationary perturbations are {\it passive}: they are written into the curvature of space during the initial epoch
and are ``discovered" as the Universe expands. As a result, they suffer from damping due to photon diffusion in the time between the initial epoch and recombination.  Strings, on the other hand, exist during all cosmological epochs, {\it actively} generating
perturbations. These perturbations come in two types:  before recombination, string wakes generate density perturbations in the primordial
plasma, leading to a single acoustic peak; after recombination, strings lens the background CMB light through the Kaiser-Stebbins effect, generating
 high-$\ell$ perturbations in the observed CMB anisotropy beyond the damping tail (for more details on string physics, see \cite{VilenkinShellardBook};  
for direct searches for string discontinuities in the
CMB maps, see \cite{Jeong:2004ut,Fraisse:2007nu,Danos:2009vv}).
String-sourced plasma perturbations before recombination are also damped by diffusion, but since strings renew those perturbations and also generate lensing-based perturbations after recombination, their perturbations are less damped.

In this paper, we use a CMB-only dataset consisting of the WMAP $7$-year data release \cite{Larson:2010gs} and the SPT $\ell < 3000$ power spectrum 
 to constrain the amplitude of a consensus zero-width (``Nambu-Goto") cosmic string spectrum using a Markov
Chain Monte Carlo (MCMC) likelihood analysis to explore constraints on its amplitude jointly with the six standard flat $\Lambda$CDM cosmological parameters. We describe our method
in more detail in \S \ref{methods}, and report the results and implications of our analysis in \S \ref{results}. Our chief results are presented in Tab.~\ref{tab:parameters_WMAP7_SPT}.  

\section{Methods}
\label{methods}
In this section we discuss the string template contribution to the CMB anisotropy spectrum and
the methodology to constrain its amplitude using the WMAP $7$-year and SPT data sets.

\subsection{String Model}
For our limits, we make use of version 3 of the publicly available code 
CMBACT \footnote{\url{http://www.sfu.ca/~levon/cmbact.html}}\cite{Pogosian:1999np}, which is based on CMBFAST \cite{Seljak:1996is}.
This code makes use of the unconnected segments model. In this model,
the full complexity of the string network is replaced with a collection of unconnected finite-length
string segments. These segments have a length, number density, and velocity distribution that evolve in time
according to the velocity-dependent one-scale model \cite{Martins:1995tg, Martins:1996jp, Martins:2000cs}. These segments are then used
to compute the string network effects on both the pre-recombination plasma and string-sourced lensing.
It is worth noting that this technique, while effective for producing two-point power spectra, cannot generate
realistic higher-point spectra nor, {\it a fortiori}, full maps of string-sourced CMB anisotropy. 

 Because of noise from finite sampling effects, 
the code averages over a large number ($N>100$) of segment collection realizations to generate smooth spectra.
The spectra it produces match those from large-scale string network
simulations (e.g. \cite{Allen:1996wi, Allen:1997ag, Landriau:2003xf, Battye:2010xz}). For our constraints, we make use of the code's standard set of string network parameters. In doing this,
we implicitly assume that the string network we will constrain is that predicted by the zero-width-approximation
Nambu-Goto string network simulations. Since string core widths are so many orders of magnitude smaller than the
string radii of curvature in the epochs relevant to the CMB, this is widely agreed to be a sound approximation.

 \begin{figure}[tb]
   \centering
   \includegraphics[width=0.5\textwidth]{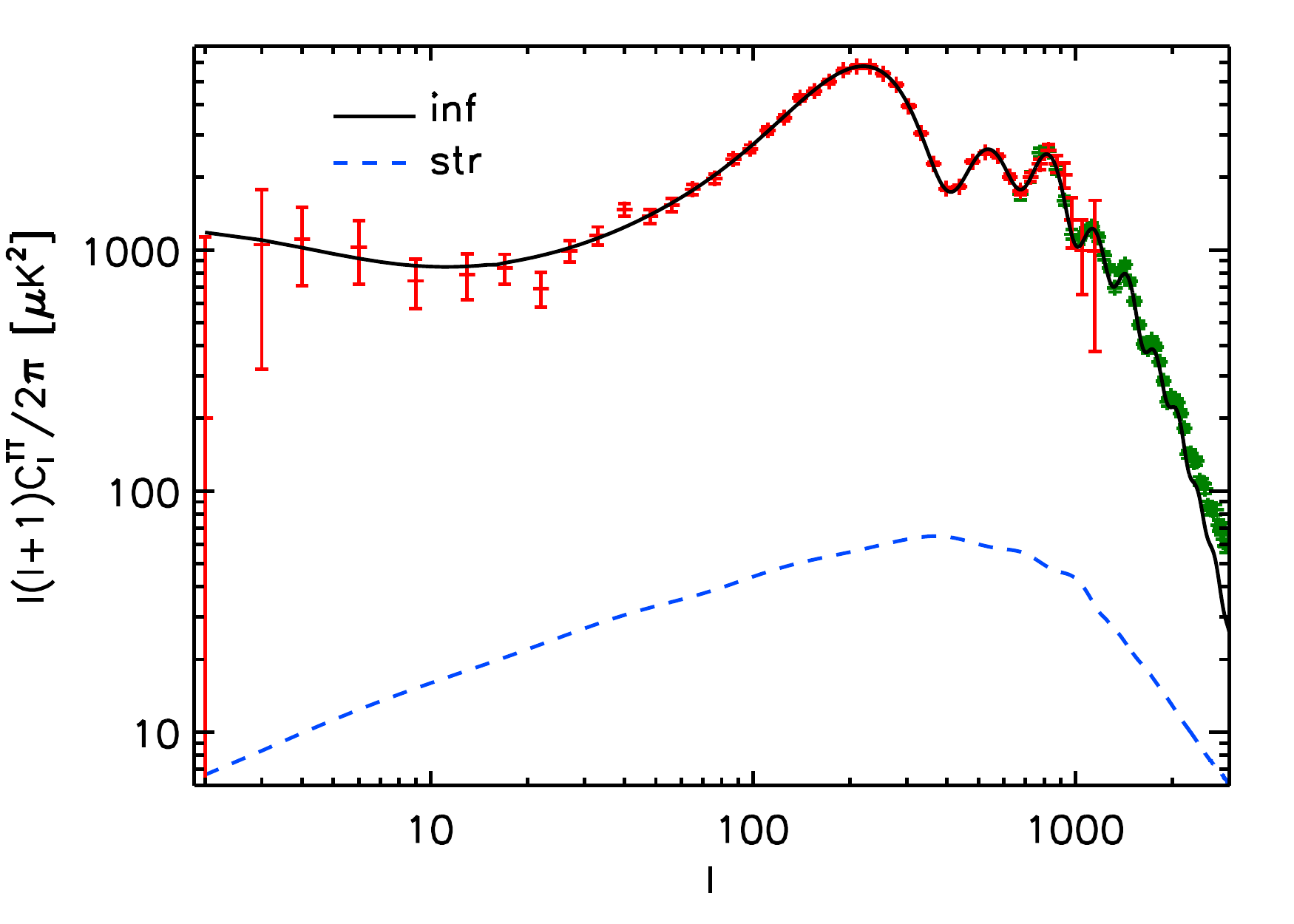}
\caption{Inflationary power spectra at its ML value from the WMAP$7$ $+$ SPT analysis in Tab. \ref{tab:parameters_WMAP7_SPT}
compared to the string template at $\fstr=0.0175$ 
($95\%$ CL limit  from WMAP$7$+SPT analysis below). The WMAP7 and SPT binned data are shown in red and green points respectively. Note that the 
plotted SPT error bars do not include beam and calibration errors; however, these errors are included in the likelihood analysis.}
\label{plot:String_vs_inflationary}
\end{figure}

The computational cost of recomputing the string spectrum for each set of cosmological
parameters is prohibitive and previous work has established that a string contribution to the
CMB anisotropy must be quite small ($<10\%$)  \cite{Albrecht:1997nt, Pogosian:2003mz, Pogosian:2004ny, Wyman:2005tu}. Thus, rather than recomputing the string spectrum for each set of cosmological parameters,
we instead compute the string spectrum only once, for the WMAP7 $+$ SPT
best fitting cosmological parameters without strings (see Tab.~\ref{tab:parameters_WMAP7_SPT}) and use
the code's default string parameters (radiation-era wiggliness $=1.05$, initial velocity $=0.4$, initial correlation length
$=0.35$) and average over $N=200$ string network realizations.
We discuss the impact of alternate choices for the string parameters and other network models in \S \ref{sec:alternate}.

  The remaining degree of freedom is then the amplitude of the string spectrum. 
We choose to normalize the template
to the  fraction of  the total CMB temperature anisotropy that can be sourced by strings 
in the small contribution limit
\begin{equation}
\fstr \equiv 
\frac{  \sigma^2_{\rm TT,str} } {\sigma^2_{\rm TT,inf}} \approx \frac{  \sigma^2_{\rm TT,str} } {\sigma^2_{\rm TT,tot}} ,
\label{fstrdef}
\end{equation}
where
\begin{equation}
\sigma_{X}^2 \equiv  \sum_{\ell =2}^{\ell_{\rm max}}  \frac{ 2\ell +1}{4\pi}  C_\ell^{X}
\end{equation}
and evaluate it with the template cosmological parameters.  For the inflationary spectrum
we again take the maximum likelihood WMAP7 $+$ SPT model from Tab.~\ref{tab:parameters_WMAP7_SPT}.
Following the previous literature \cite{Wyman:2005tu}, we take $\ell_{\rm max}=2000$
so as to reflect the fraction of power in the main acoustic peaks rather than the damping tail.
With these conventions and the fiducial string parameters, the string tension is related to $\fstr$ as
\begin{equation}
G\mu =  1.27\times10^{-6} \fstr^{1/2}.
\label{eqn:conversion}
\end{equation}

We show the string template compared with the inflationary spectrum in Fig.~\ref{plot:String_vs_inflationary}.  Here we have taken $\fstr = 0.0175$ motivated by the WMAP7 and SPT joint analysis below.  For comparison, we also plot their respective power spectrum
measurements.   Note that at $\ell > 2000$, the SPT power spectrum has a substantial
contribution from foregrounds as we shall discuss below.

\begin{figure}[tb]
   \centering
   \includegraphics[width=0.45\textwidth]{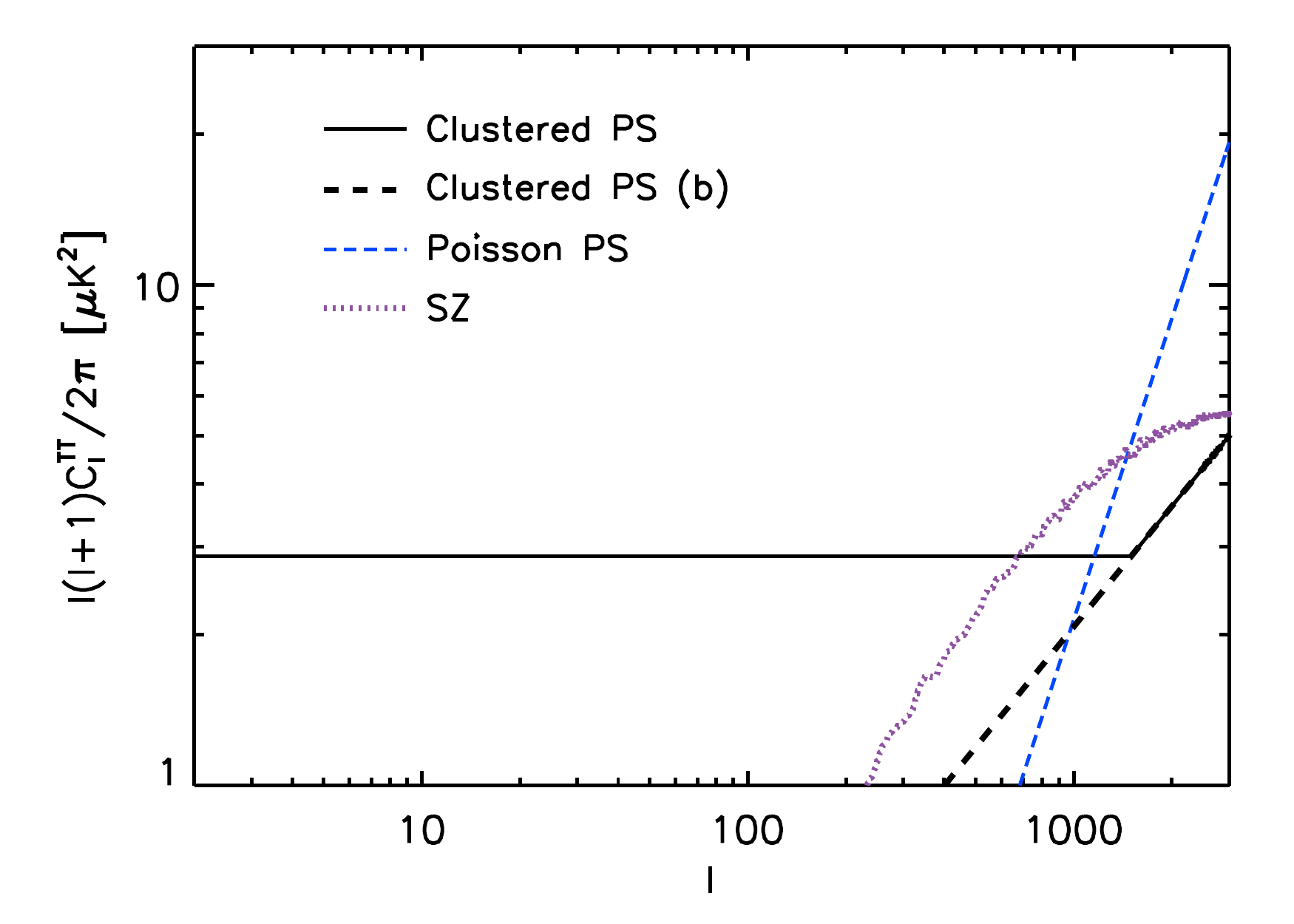}
\caption{SPT foreground templates (SZ, clustered PS, and Poisson point source spectra) shown at the central values of their priors (from \cite{Keisler:2011aw}).
Also shown is an alternate power law extension template (b) for clustered PS; we used this alternate template to test the robustness of our results to 
foreground assumptions.}
\label{plot:Foregrounds_MLCL}
\end{figure}
 
\subsection{Likelihood Analysis}

Using the six flat $\Lambda$CDM cosmological parameters and $\fstr$, we conduct two
 Markov Chain Monte Carlo (MCMC) likelihood analyses: one using only the WMAP7 data and another
combining the WMAP and SPT data sets.  We assume non-informative, flat priors on each of the 7 parameters
\begin{equation}
\{ \fstr, \Omega_b h^2, \Omega_c h^2, \theta, \tau, A_s, n_s \},
\end{equation}
where
$\Omega_b h^2$ and $\Omega_c h^2$ give the physical baryon and cold dark matter densities respectively,
 $\theta$ is 100 times the angular size of the sound horizon at recombination
and depends on the Hubble constant $h$, $\tau$ is the optical depth through reionization, 
and the inflationary curvature power spectrum is
given by $k^3 P_{\cal R}/2\pi^2=A_s (k/0.05{\rm Mpc}^{-1})^{n_s-1}$.

For the joint analysis, we additionally  marginalize over the amplitudes of three SPT foreground contaminants, using priors 
produced by a previous multifrequency analysis \cite{Shirokoff:2010cs},
as described in \cite{Keisler:2011aw}.   More specifically: we
impose Gaussian priors on the amplitude of their contribution at $\ell=3000$ (defining 
$d_\ell \equiv \ell(\ell+1)C_\ell/2\pi$): 
$d_{3000}^{\rm SZ}=5.5\pm3.0$ $\mu$K$^2$, $d_{3000}^{\rm Poisson}=19.3\pm3.0$ $\mu$K$^2$, and $d_{3000}^{\rm clustered}=5.0\pm2.5$ $\mu$K$^2$ for the Sunyaev-Zeldovich (SZ), Poisson and clustered point source components respectively.  The smallness of these amplitudes
implies that there is relatively little room for string sourced power at high-$\ell$. Even before running our full analysis, we can infer that
at $\ell=3000$ the contributions allowed from strings will be limited to being less than a few
$\mu$K$^2$, which translates into a constraint on $\fstr$ of order a percent -- very close to the result from our full analysis (see Tab. \ref{tab:parameters_WMAP7_SPT}). 

Combining these 3 amplitude parameters with
their respective power spectrum templates (shown in  Fig. \ref{plot:Foregrounds_MLCL}), 
we reproduce the full $d_\ell$ (foreground) model used in the SPT likelihood function.    
Following \cite{Keisler:2011aw}, we 
combine the WMAP7 and SPT CMB likelihoods as if they were completely independent.   Given
that SPT covers only 790 sq. degrees of sky and only significantly augments WMAP for an
$\ell$-range where WMAP7 is noise limited, this is a good approximation.  
The SZ amplitude prior from SPT limits the possible SZ contamination to WMAP7 to be negligible at its central value; hence, 
we do not include SZ as a foreground parameter for WMAP7 \cite{Keisler:2011aw}.

\begin{figure}[tbp]
   \centering
   \includegraphics[width=0.45\textwidth]{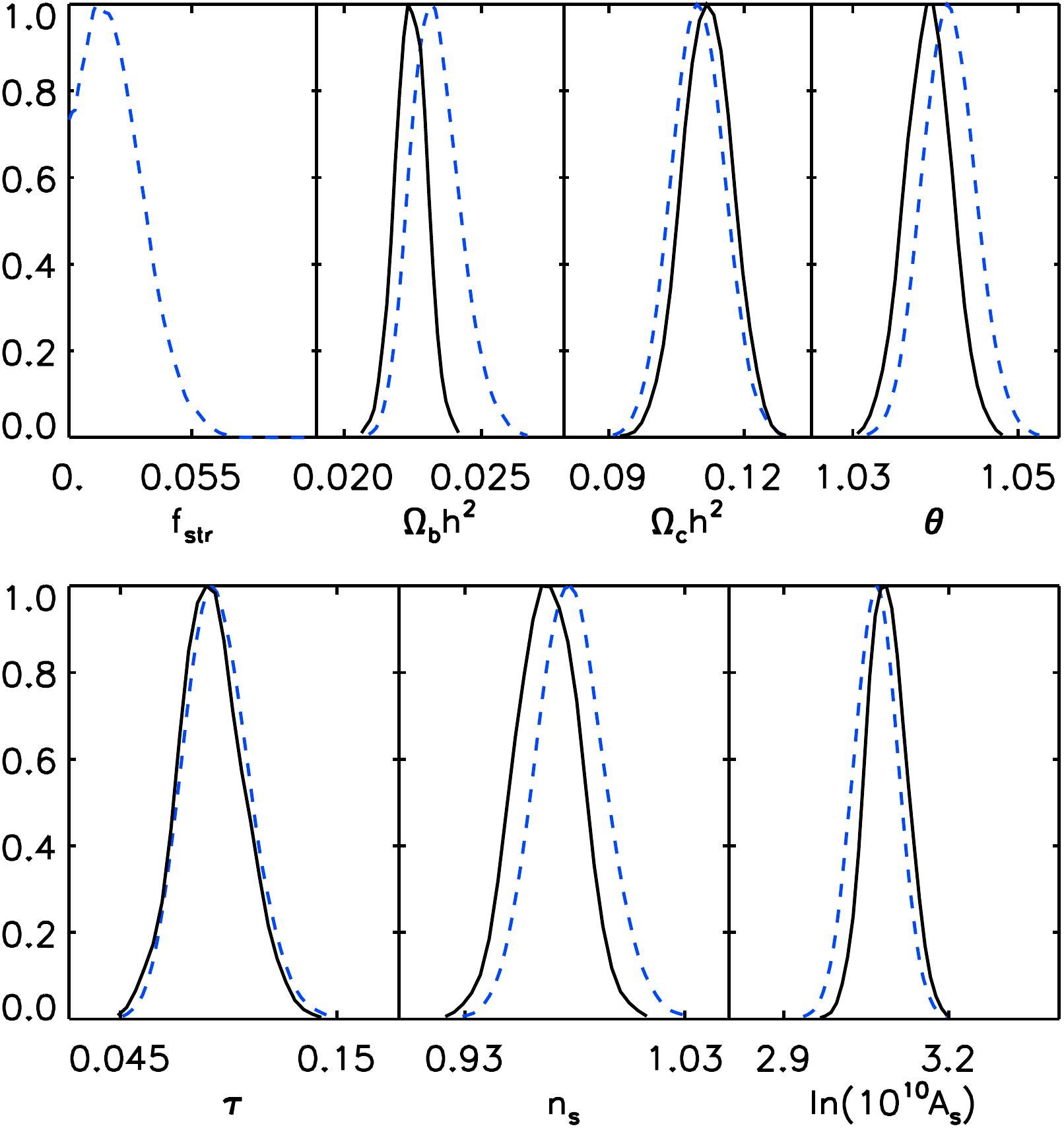}
	\caption{Parameter probability distributions with WMAP$7$ data only in a flat universe. Blue/dashed lines represent the posteriors when considering the string contribution, and black lines represent the posteriors without the string contribution. } 
   \label{plot:WMAP7_with_without_Strings}
\end{figure}

\begin{figure}[tbp]
   \centering
   \includegraphics[width=0.45\textwidth]{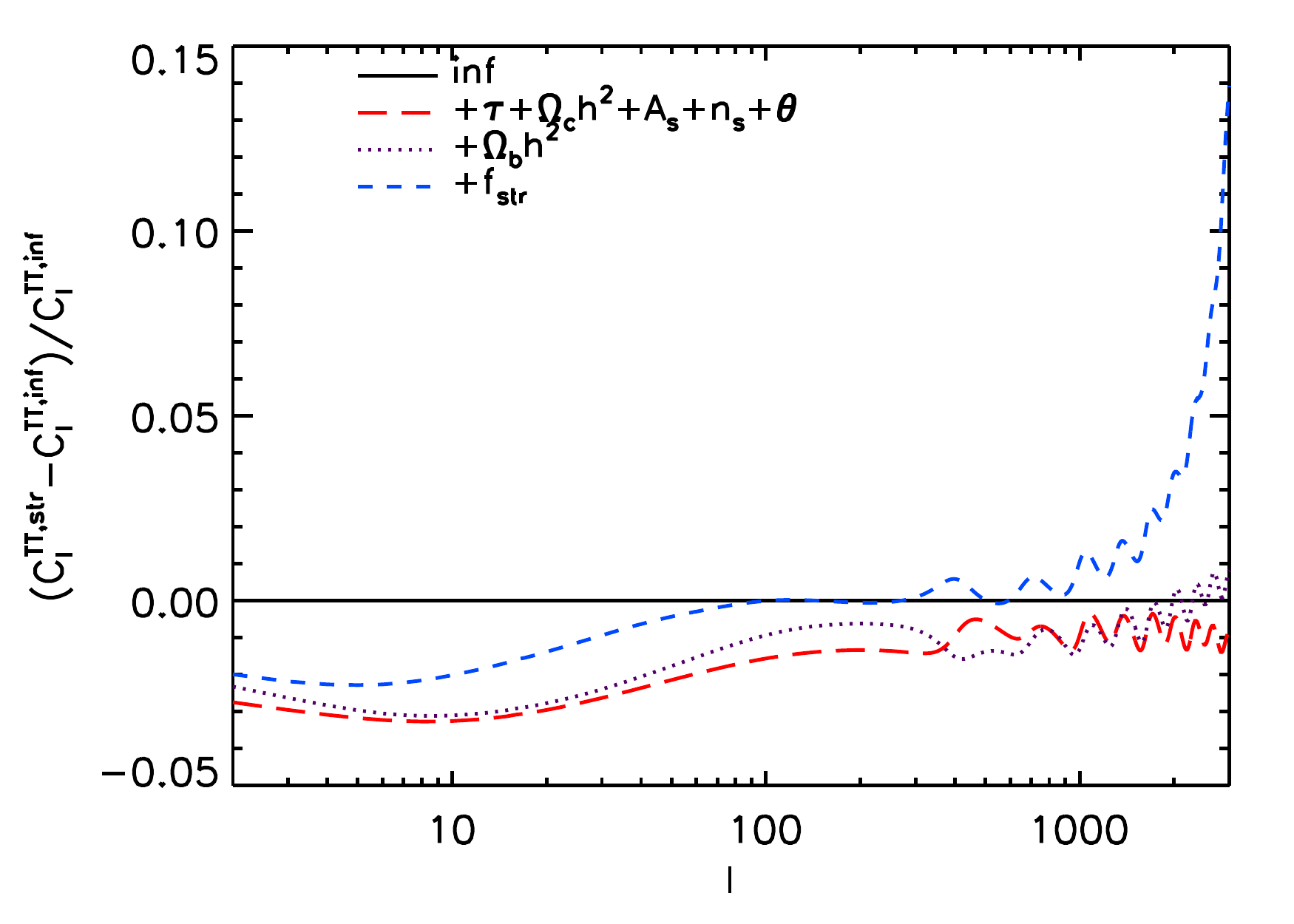}
\caption{Approximate degeneracy between strings and baryons in $C_\ell^{\rm TT}$
at $\ell < 1000$.    Shown is 
the  fractional difference between the maximum likelihood inflation-only power spectrum and a series of alternate spectra
where the listed parameters are cumulatively changed from their values in this model to those of
 the inflation $+$ $\fstr$ maximum likelihood model using {\it only} WMAP$7$ data.  Raising  the baryon density compensates the contribution from strings -- which peaks near the second acoustic peak -- leaving the spectrum for $\ell < 1000$ nearly 
unchanged.   That the black line is close to the blue/dashed line here shows why $\fstr < 0.046$ is allowed by WMAP alone. The rapid rise of the blue/dashed line for $\ell>1000$, in turn, explains why SPT is able so tightly
to constrain $\fstr$ ({\it cf.}~\cite{Pogosian:2008am}).}
   \label{plot:DeltaClTT_fromLCDMtoGmu}
\end{figure}

\section{Results} 
\label{results}

In this section we present our main results from the likelihood analysis of the string fraction 
with the WMAP7 and SPT 
data sets as well as  discuss their implications for future polarization experiments and caveats associated with alternate string models.
 
\subsection{Temperature Constraints}

The dominant constraint from WMAP7 and SPT comes from their measurement of the
temperature anisotropy spectrum.    In order to establish a baseline for comparison,
we begin with an analysis of constraints from WMAP7 alone.  
In Fig.~\ref{plot:WMAP7_with_without_Strings}, we show the posterior probability distribution of the parameters with and without $\fstr$.   
Note that with $\fstr$ the $\Omega_b h^2$ 
distribution widens toward higher values and the peak of the $\fstr$ probability distribution is at a positive value \cite{Pogosian:2003mz}.

Tab. \ref{tab:parameters_WMAP7_SPT} shows the means, standard deviations and maximum likelihood values of the WMAP7 only analysis (with and without the contribution from strings), compared with the analysis of the WMAP$7$ $+$ SPT datasets.
For the string contribution, there is no detection at $95\%$ CL so we instead quote the
1 sided $95\%$ upper limit $\fstr < 0.0457$.

\begin{table*}[tbp]
\centering
\begin{tabular}{|c|cccc|cccc|} 
\hline
Parameters & \multicolumn{4}{c|}{WMAP7} & \multicolumn{4}{c|}{WMAP7$+$SPT}\\ 
\hline
& \multicolumn{2}{c|}{Without $\fstr$} &  \multicolumn{2}{c|}{With $\fstr$} &  \multicolumn{2}{c|}{Without $\fstr$} &  \multicolumn{2}{c|}{With $\fstr$} \\ 
\hline
$\fstr$ & \multicolumn{1}{c|}{-} &  \multicolumn{1}{c|}{-} &  \multicolumn{1}{c|}{$<0.0457$} &  \multicolumn{1}{c|}{$0.0101$} &  \multicolumn{1}{c|}{-} &  \multicolumn{1}{c|}{-} &  \multicolumn{1}{c|}{$<0.0175$} &  \multicolumn{1}{c|}{$0.0030$} \\ 
\hline
$100\Omega_bh^2$ & \multicolumn{1}{c|}{$2.243\pm0.056$} &  \multicolumn{1}{c|}{$2.237$} &  \multicolumn{1}{c|}{$2.335\pm0.089$} &  \multicolumn{1}{c|}{$2.286$} &  \multicolumn{1}{c|}{$2.222\pm0.041$} &  \multicolumn{1}{c|}{$2.218$} &  \multicolumn{1}{c|}{$2.239\pm0.045$} &  \multicolumn{1}{c|}{$2.228$} \\ 
\hline
$\Omega_ch^2$ & \multicolumn{1}{c|}{$0.1118\pm0.0053$} &  \multicolumn{1}{c|}{$0.1122$} &  \multicolumn{1}{c|}{$0.1098\pm0.0059$} &  \multicolumn{1}{c|}{$0.1108$} &  \multicolumn{1}{c|}{ $0.1117\pm 0.0048$} &  \multicolumn{1}{c|}{$0.1121$} &  \multicolumn{1}{c|}{$0.1105\pm0.0049$} &  \multicolumn{1}{c|}{$0.1114$} \\ 
\hline
$\theta$ & \multicolumn{1}{c|}{$1.0392\pm0.0028$} &  \multicolumn{1}{c|}{$1.0389$} &  \multicolumn{1}{c|}{$1.0417\pm0.0032$} &  \multicolumn{1}{c|}{$1.0400$} &  \multicolumn{1}{c|}{$1.0410\pm0.0016$} &  \multicolumn{1}{c|}{$1.0412$} &  \multicolumn{1}{c|}{$1.0413\pm0.0016$} &  \multicolumn{1}{c|}{$1.0408$} \\ 
\hline
$\tau$ & \multicolumn{1}{c|}{$0.089\pm0.015$} &  \multicolumn{1}{c|}{$0.089$} &  \multicolumn{1}{c|}{$0.092\pm0.016$} &  \multicolumn{1}{c|}{$0.089$} &  \multicolumn{1}{c|}{$0.085\pm0.014$} &  \multicolumn{1}{c|}{$0.087$} &  \multicolumn{1}{c|}{$0.085\pm0.014$} &  \multicolumn{1}{c|}{$0.083$} \\ 
\hline
$n_s$ & \multicolumn{1}{c|}{$0.9674\pm0.0139$} &  \multicolumn{1}{c|}{$0.9668$} &  \multicolumn{1}{c|}{$0.9780\pm0.0156$} &  \multicolumn{1}{c|}{$0.9742$} &  \multicolumn{1}{c|}{$0.9652\pm0.0109$} &  \multicolumn{1}{c|}{$0.9632$} &  \multicolumn{1}{c|}{$0.9646\pm0.0112$} &  \multicolumn{1}{c|}{$0.9634$} \\ 
\hline
$\ln[10^{10}A_s]$ & \multicolumn{1}{c|}{$3.0860\pm0.0349$} &  \multicolumn{1}{c|}{$3.0879$} &  \multicolumn{1}{c|}{$3.0665\pm0.0399$} &  \multicolumn{1}{c|}{$3.0762$} &  \multicolumn{1}{c|}{$3.0806\pm0.0300$} &  \multicolumn{1}{c|}{$3.0849$} &  \multicolumn{1}{c|}{$3.0667\pm0.0324$} &  \multicolumn{1}{c|}{$3.0689$} \\ 
\hline
$H_0$ & \multicolumn{1}{c|}{$70.42\pm2.38$} &  \multicolumn{1}{c|}{$70.08$} &  \multicolumn{1}{c|}{$72.91\pm3.19$} &  \multicolumn{1}{c|}{$71.44$} &  \multicolumn{1}{c|}{$70.97\pm2.16$} &  \multicolumn{1}{c|}{ $70.77$} &  \multicolumn{1}{c|}{$71.66\pm2.28$} &  \multicolumn{1}{c|}{$71.02$} \\ 
\hline
$\Omega_\Lambda$ & \multicolumn{1}{c|}{$0.728\pm0.028$} &  \multicolumn{1}{c|}{$0.726$} &  \multicolumn{1}{c|}{$0.747\pm0.031$} &  \multicolumn{1}{c|}{$0.738$} &  \multicolumn{1}{c|}{$0.733\pm0.025$} &  \multicolumn{1}{c|}{$0.732$} &  \multicolumn{1}{c|}{$0.740\pm0.025$} &  \multicolumn{1}{c|}{$0.735$} \\ 
\hline
$d_{3000}^{\rm SZ}$ & \multicolumn{1}{c|}{-} &  \multicolumn{1}{c|}{-} &  \multicolumn{1}{c|}{-} &  \multicolumn{1}{c|}{-} &  \multicolumn{1}{c|}{$5.16\pm2.27$} &  \multicolumn{1}{c|}{$4.93$} &  \multicolumn{1}{c|}{$4.30\pm2.21$} &  \multicolumn{1}{c|}{$5.43$} \\ 
\hline
$d_{3000}^{\rm Poisson}$ & \multicolumn{1}{c|}{-} &  \multicolumn{1}{c|}{-} &  \multicolumn{1}{c|}{-} &  \multicolumn{1}{c|}{-} &  \multicolumn{1}{c|}{$20.33\pm2.37$} &  \multicolumn{1}{c|}{$21.35$} &  \multicolumn{1}{c|}{$19.61\pm2.45$} &  \multicolumn{1}{c|}{$19.87$} \\ 
\hline
$d_{3000}^{\rm clustered}$ & \multicolumn{1}{c|}{-} &  \multicolumn{1}{c|}{-} &  \multicolumn{1}{c|}{-} &  \multicolumn{1}{c|}{-} &  \multicolumn{1}{c|}{$5.23\pm2.12$} &  \multicolumn{1}{c|}{$4.79$} &  \multicolumn{1}{c|}{$4.69\pm2.08$} &  \multicolumn{1}{c|}{$4.77$} \\ 
\hline
-2$\ln L$ & \multicolumn{2}{c|}{$7468.8$} &  \multicolumn{2}{c|}{$7468.1$} &  \multicolumn{2}{c|}{$7506.7$} &  \multicolumn{2}{c|}{$7506.6$} \\ 
\hline
\end{tabular}
\caption{Means, standard deviations and maximum likelihood values of the WMAP7 data analysis (with and without the contribution from strings), compared with the analysis of the WMAP$7$ $+$ SPT data sets. The upper limits for $\fstr$ refer to the $95\%$ CL limit.}
\label{tab:parameters_WMAP7_SPT}
\end{table*}

The upper limit from WMAP7 alone on $\fstr$ is weakened by a 
degeneracy between strings and baryons.   In
Fig.~\ref{plot:DeltaClTT_fromLCDMtoGmu}, we show the impact of changing sets of
parameters from the maximum likelihood model without to with strings.   Note that the string spectrum peaks around the second
acoustic peak in the CMB, at $\ell \sim 550$.   Increasing the baryon density decreases
the second peak and helps compensate the additional power from strings.   
The net result of these parameter variations is that parameter degeneracies make the two models nearly indistinguishable in the
WMAP range of $\ell < 1000$ out to $\fstr \sim 0.05$.  

\begin{figure}[tbp] 
   \centering
   \includegraphics[width=0.45\textwidth]{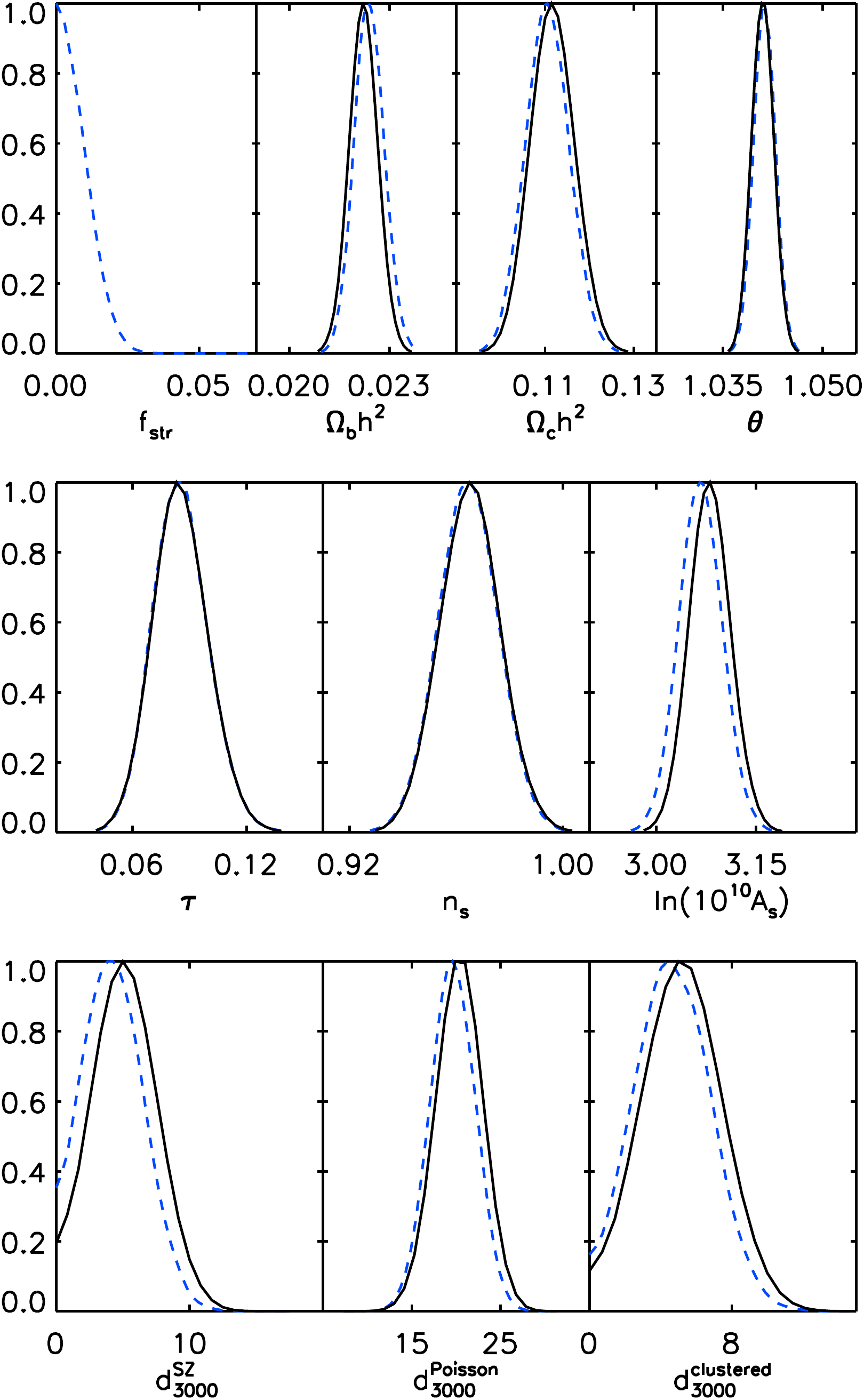}
   \caption{Parameter probability distributions with WMAP$7$ $+$ SPT data in  a flat universe. Blue/dashed lines represent the posteriors when considering the string contribution, and black lines represent the posteriors without the string contribution. }
   \label{plot:SPT_with_without_Strings}
\end{figure}

For $\ell> 1000$ the accidental parameter degeneracy is broken.    String contributions
become a larger fraction of the total and cannot be compensated by adjusting the
standard cosmological parameters.    Thus the addition of SPT data should be able to improve the constraint on $\fstr$ by more than just the additional
statistical power of the joint data set ({\it cf.}~\cite{Pogosian:2008am}).

\begin{figure}[tbp] 
   \centering
   \includegraphics[width=0.45\textwidth]{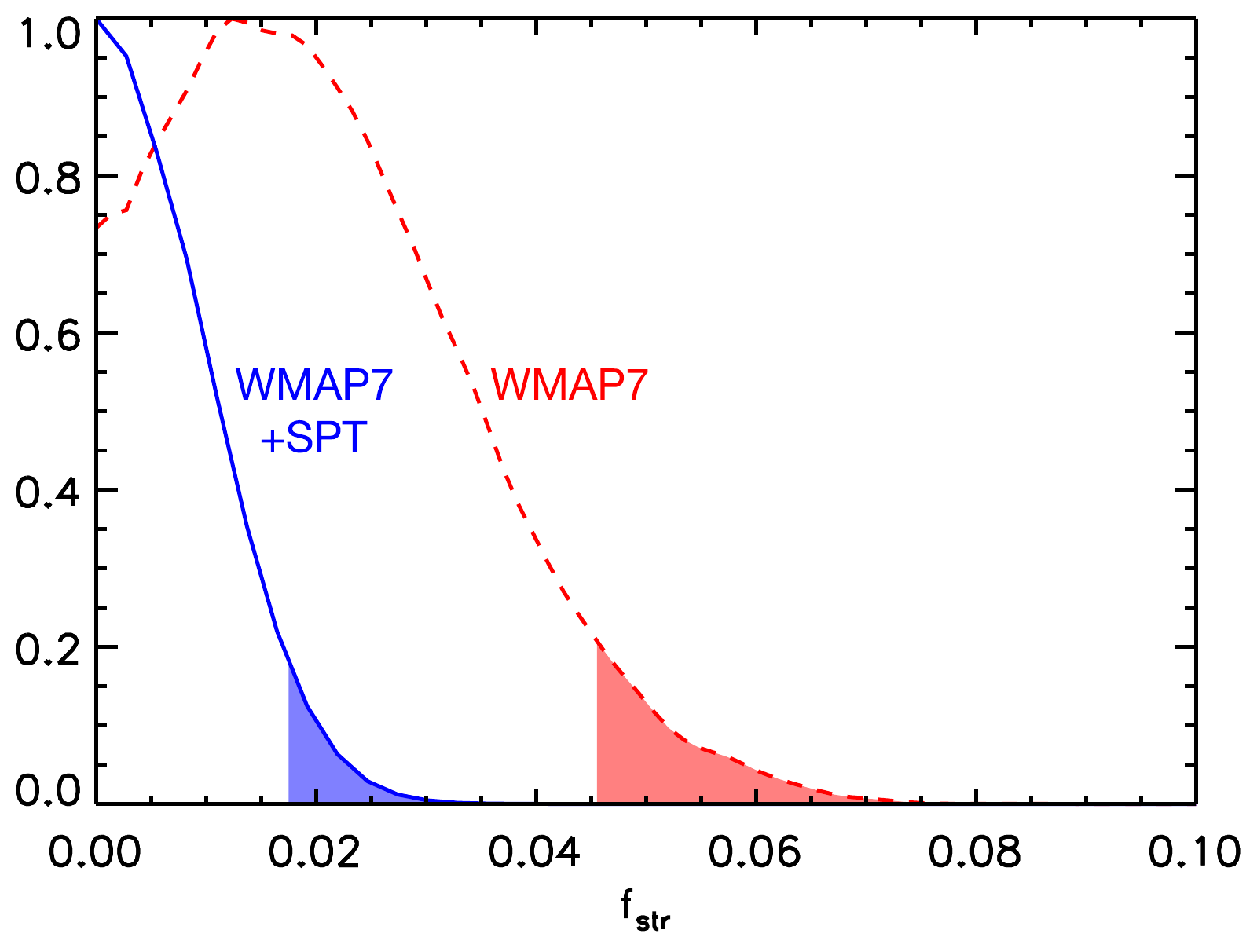} 
   \caption{The posterior probability of the parameter $\fstr$ for WMAP7 (in red/dashed lines) alone and WMAP7 + SPT (in blue/solid lines). The shaded region corresponds to the area with $\fstr$ greater than its $95\%$ CL limit value.}
   \label{plot:fstrposterior}
\end{figure}

Indeed, the joint WMAP7 and SPT analysis shows this improvement.   In Fig.~\ref{plot:SPT_with_without_Strings}, we show the posterior probability distributions for the
full 10 parameter set including SPT foregrounds.   The addition of $\fstr$ now no longer
significantly affects the other parameters, indicating little remaining degeneracy.  Furthermore
the $\fstr$ distribution peaks at zero.    As shown in Fig.~\ref{plot:fstrposterior},
the result is that the 95\% CL upper limit
improves by a factor of 2.6 to $\fstr <0.0175$  or using Eq.~(\ref{eqn:conversion}), $G\mu < 1.7 \times 10^{-7}$.

Since this limit is close to the foreground uncertainty limit at $\ell\sim 3000$, it is important
to check its robustness to foreground assumptions.   Given the limited frequency spectrum
separation of SZ and blackbody contributions and the downward shift in the posterior 
of $d_{3000}^{\rm SZ}$  when the string contribution is marginalized, we first consider removing the Gaussian SZ prior and replacing it with a flat uninformative one.  The result is a very small
weakening of the limit to $\fstr  < 0.0184$.   
Next, we consider modifications in the SPT clustered point source template.   
As shown in Fig.~\ref{plot:Foregrounds_MLCL}, the default template was arbitrarily set to a constant at low $\ell$.
We replace the constant with a power law extension of  $d_\ell^{\rm clustered} \propto \ell^{0.79}$, consistent with the
$d_\ell^{\rm clustered} \propto \ell^{0.75\pm 0.06}$ finding of  \cite{Addison:2011se}, that
smoothly joins onto the high $\ell$ template.
Again there is a negligible impact on the string constraint $\fstr<0.0173$.  We conclude that our results are robust to at least mild changes in the foreground modeling.

\begin{figure}[tbp] 
   \centering
   \includegraphics[width=0.5\textwidth]{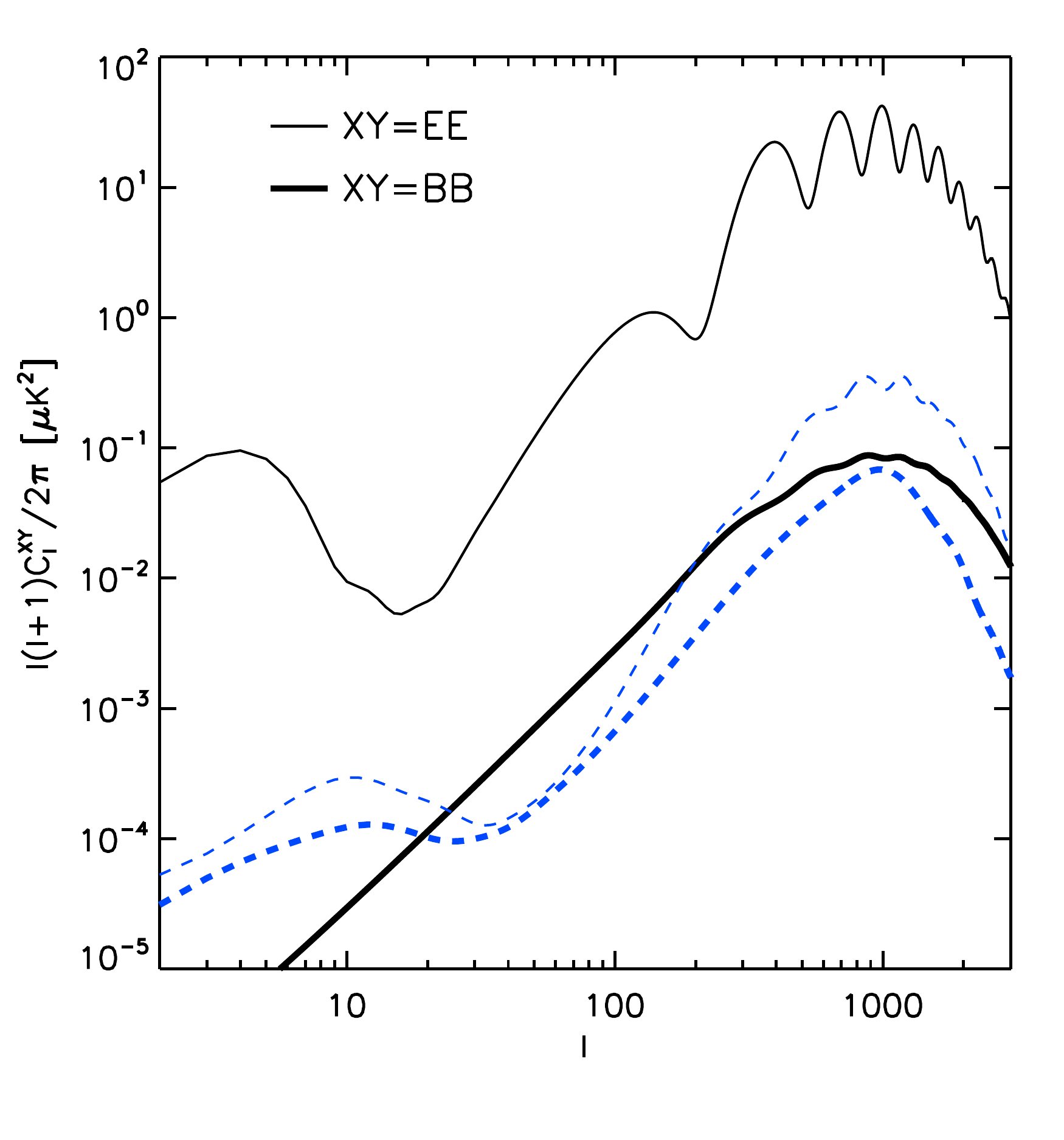} 
   \caption{Polarization power spectra for the inflationary template (in black/solid lines) compared with the maximal contribution from strings that satisfies the WMAP7 $+$ SPT constraint  at the $95\%$CL limit (in blue/dashed lines).   EE (BB) spectra are plotted with
   thin (thick) lines.  BB string contributions of order or greater than the lensing contribution to the inflationary spectrum are still allowed but only at $\ell \sim 1000$ and $\ell \lesssim 10$.}
   \label{plot:polarization}
\end{figure}

\subsection{Polarization Prospects}

While current polarization measurements do not contribute substantially to the constraint on 
the string fraction $\fstr$, it is interesting to ask, in light of the improved temperature
constraint, how polarization measurements may improve future results.
In this section we study the polarization prospects for constraining the string fraction. This subject has a long literature (beginning with e.g. \cite{Seljak:1997ii, Battye:1998js, Benabed:1999wn}). The main goal of this section is to point out  that the best hope for detecting cosmic strings will come from arcminute measurements of the B-modes (pursued by experiments seeking to measure the gravitational lensing of the CMB) rather than from degree scales (pursued by experiments looking for gravitational waves). This is a consequence of the bound on the cosmic string tension coming from the temperature spectrum.

Fig.~\ref{plot:polarization} shows the predicted polarization spectra for a model
with $\fstr$ at the current 95\% CL bound.    Note that the fraction of the string contribution to
the EE polarization is similar but smaller
\begin{equation}
\fstr^{\rm EE} \approx 0.55 \fstr
\end{equation}
(where $\fstr^{\rm EE}$ is defined as in Eq.~(\ref{fstrdef}) with the replacement TT $\rightarrow$ EE).
This implies that generally EE polarization will only begin to assist constraints when
measurements become comparably accurate as TT.    This may occur due to improvements
in sensitivity or due to difficulties in foreground subtraction for TT at high $\ell$. 

On the other hand, the EE polarization fraction is smaller mainly due to the comparable
contribution to BB from strings.   Moreover for the inflationary spectrum the BB total
power is dominated by gravitational lensing conversion of E-mode polarization (rather than primary anisotropy); hence the
fraction that strings contribute is substantially higher for BB:
\begin{equation}
\fstr^{\rm BB} \approx 32 \fstr.
\end{equation}
Therefore as soon as polarization begins to help constraints, the dominant improvement on string constraints
will come from the B modes.

The current 95\% CL bound on $\fstr$ allows a string contribution to BB that is of the same order as the gravitational
lensing BB signature on similar angular scales peaking at $\ell \sim 1000$. Current B-mode experiments put bounds \cite{Ma:2010yb} that are not competitive with temperature based bounds, but in the future, experiments that are designed to have the high angular resolution and precision to measure gravitational lensing B-modes, {\it e.g.}\ SPTpol and ACTpol, will thus also either improve the
string constraint, or possibly detect their contribution.

Conversely, experiments that seek to detect inflationary gravitational waves in B modes
at $\ell \lesssim 100$ are unlikely to improve string bounds significantly.    The upper bound on $\fstr$ from the temperature spectrum already implies that the string contributions to the BB power spectrum are smaller than lensing at $\ell = 100$ by a factor of $4$. While reionization creates BB string fluctuations at $\ell \lesssim 10$, these are constrained to
be of very small amplitude. 
These qualitative arguments are
confirmed by more detailed studies performed in, {\it e.g.} \cite{Bevis:2007qz, Mukherjee:2010ve, Avgoustidis:2011ax}.

\subsection{Alternate Models and Other Considerations}
\label{sec:alternate}

In this paper, we have selected a fixed cosmic string spectrum to constrain: the spectrum generated by a single species of local
cosmic strings in the Nambu-Goto or zero-width approximation using consensus network parameters.  We should mention, however,
that there are a number of alternate spectra that we could also have studied. These alternate spectra are motivated in two ways: 1) when we deviate
from the simple local, single-species model and 2) when simulations are performed using finite-width or ``Abelian Higgs" network assumptions. Let us 
discuss these in turn.

A network of local cosmic strings where all strings have the same tension is the classic and by far the best studied example of
a cosmic string network. Nonetheless, there are other models that are worth consideration. Perhaps the simplest modification would be to replace local strings
with semilocal strings \cite{Achucarro:1999it}. Whereas local strings are absolutely stable thanks
to conservation of topological charge, semilocal strings are only metastable -- they can decay ({\it n.b.}~even local strings can decay in string theory models, albeit in a different manner; {\it cf.}~\cite{Myers:1900zz}). As a practical matter, this leads to a  different
spectrum of perturbations \cite{Urrestilla:2007sf}. 

Another possibility is that multiple types of cosmic strings exist. 
This occurs in string theory \cite{Copeland:2003bj}, where multiple string types and an infinite tower of bound states can arise; 
although in practice only the lowest tension species tend to dominate
the network dynamics \cite{Tye:2005fn}. Despite this, realistic multi-tension / multi-species networks can produce distinctive string
spectra \cite{Avgoustidis:2011ax}. Another string theory effect that we neglect is reduced intercommutation probabilities between strings,
which can lead to higher string number densities \cite{Jackson:2004zg}, though this effect has been shown to be more muted in realistic
simulations than simple analytical estimates suggest \cite{Avgoustidis:2005nv}.

All stringy spectra are qualitatively similar -- a single broad peak and slower-than-inflationary fall-off at high-$\ell$ -- but
they do differ in details. The most important such detail is the location and sharpness of the principle peak. The effective
correlation length and rms velocity of the strings alter this: slower strings give a higher peak, while longer (shorter)
effective correlation lengths lead to a lower (higher) peak location in $\ell$ \cite{Pogosian:2007gi}. 
 Other more subtle effects can also alter the peak's width \cite{Mukherjee:2010ve, Foreman:2011uj}. 
 A strong prediction
for this peak's shape and location would make it easier to detect on top of the expected lensing of E-mode to B-mode polarization, 
but the theoretical uncertainty introduced by the existence of these alternate models disallows such a definite prediction. 

All the considerations mentioned in the preceding paragraph presuppose calculations that rely on zero-width string network calculations. 
Although this is widely regarded as a valid approximation for calculating cosmic string network effects, some dissent from this viewpoint.
In particular, simulations performed by directly solving the equations
of motion that describe the fields that constitute the strings produces quite different predictions from those found
in the zero-width simulations \cite{Bevis:2006mj}. The CMB spectrum inferred from the finite-width simulations has a broader,
lower amplitude peak that lies at larger angular scales for fixed string tension as compared with the spectrum from the zero-width simulations. This difference
comes about because the finite-width simulations turn string ``wiggliness" (high frequency oscillations generated by
string-string collisions) into particle production,  whereas the zero-width simulations assume that ``wiggliness" is damped only 
by gravitational physics that operates too slowly to make a significant impact network behavior. 

The contradiction between
these results has never definitively been resolved. However, it is difficult to know how far to trust these results when the simulations must be
performed on the extremely small length scales necessary to resolve string cores. The results must be extrapolated by tens of orders of magnitude to 
connect the core-scale physics to the cosmological length scales where string effects are manifested in the CMB. 
Despite the careful efforts that have been made to check the consistency of all aspects of the simulations \cite{Bevis:2010gj}, the techniques necessary
to resolve string cores -- which include the limitations of small  dynamical ranges ($\leq1024^3$ grids) and 
the introduction of artificially expanding cores in some (though not all) simulations -- are precisely those that emphasize finite-width effects.
It is furthermore worth mentioning that even if the finite-width simulations accurately represent the cosmological behavior of strings that are described
by an Abelian Higgs-type theory, cosmic strings that come from string theory would likely have microscopic physics of a very different character \cite{Polchinski:2004ia, Jackson:2004zg, Myers:1900zz}
which would be less likely to permit wiggliness to stimulate the emission of quanta.
In light of these concerns, we do not study the finite-width string spectra in this paper \footnote{N.B., beyond these theoretical concerns,
the finite-width spectra also have significantly less power at high-$\ell$ as compared with zero-width spectra \cite{Battye:2010xz, Bevis:2007gh}.
This will lead to less significant improvement to the constraints on strings as compared with what we find; see a recent and closely related work, \cite{Urrestilla:2011gr}}. 

In fairness, we should make clear that the results that we rely on are also 
subject to possibly dangerous extrapolation effects. We rely on an unconnected
segments model to approximate the behavior of a zero-width string network. Neither step can be definitively proven to be the optimal method
for analyzing the large scale behavior of string networks.
For instance, assuming that string cores have no effect on network behavior is baked
in to the zero width approach to string network modeling and hence is not tested by that modeling. 
Similarly, the unconnected segments model has parameters that must be fitted to simulations, hence is also subject 
to modeling errors. We could, in principle, simply allow these parameters to vary in our analysis.
Allowing the parameters to vary will permit the string spectrum's quantitative features (peak location, height, high-$\ell$ fall-off) to 
change, too, which would likely lead to weaker constraints on the string tension \cite{Foreman:2011uj}. In doing this we would give
up the a priori predictivity of our method in favor of the hope that the data themselves could give information about the string network.
If or when there is some observational evidence for a stringy contribution to the CMB, such a
parameter search would become a very interesting thing to perform.

We also note that pulsar timing \cite{Jenet:2006sv,vanHaasteren:2011ni} provides a constraint on string networks that is complementary to CMB-based tests like ours. This
is because cosmic string networks provide a source for the stochastic gravity wave background. Unlike CMB based constraints, which are
directly sensitive to the string tension, $G\mu$, pulsar timing measurements limit a combination of the string tension and the effective
string loop formation length scale, $\alpha_{\rm eff}$: $\Omega_{\rm GW} \propto (\alpha_{\rm eff} G\mu)^{1/2}$ \cite{Myers:1900zz, Polchinski:2007qc}.
The value $\alpha_{\rm eff}$ is highly model dependent. Different techniques give estimates for the typical loop size that differ from each other
by many orders of magnitude. The upshot of this is that pulsar timing constraints must be treated with caution, whereas CMB constraints like ours are
considerably more robust.  
All that being said, when calculated using a combination of some of the current best analytical and numerical techniques,
the resulting pulsar timing constraint is $G\mu \lesssim 2\times10^{-7}$  \cite{Polchinski:2007qc}, similar to the constraint we find. See \cite{Battye:2010xz, Polchinski:2007qc} for further discussion and many more references.
We elected to exclude non-CMB datasets to give a clean and simple determination of what the CMB alone could say about cosmic string networks.

\section{Conclusion}
By combining the observations of the CMB made by the WMAP and SPT experiments, we are able
to produce the strongest constraints yet obtained on the tension of a network of cosmic strings:
$G\mu <1.7 \times10^{-7}$ at 95\% confidence. This corresponds to a limit on the CMB
power produced by strings below $\ell=2000$ of $\fstr < 0.0175$.     
While in principle future temperature based constraints can push this limit an order of magnitude lower, in practice
foreground modeling will eventually place a systematic floor on temperature-based limits. Hence the limit
we report is likely within a factor of a few of the best obtainable limit on strings using only measurements of the CMB temperature anisotropy.
Our limit is robust to uncertainties in the amplitude of the SZ contamination as well as power law clustered point source contributions.

In addition to providing a new set of bounds for comparison in model building, these limits also
define what future experiments will be needed either to discover or to yet more tightly constrain strings. 
Polarization measurements have long been advocated as a means of improving these
constraints further since the fractional contribution from strings to the B-modes is much higher
than in temperature.      
Unfortunately, given our constraint, large-angle polarization experiments
designed to detect inflationary gravitational waves are unlikely to improve constraints on 
zero-width Nambu-Goto strings.   Their contribution is already limited
to be a small fraction of the lensing B-modes at $\ell < 100$.    The opportunity to improve
constraints lies in the $\ell \sim 1000$ range, where the bounds we report still permit
strings to make an order unity contribution to B-mode polarization as compared with the
B-modes expected from gravitational lensing of inflationary E-modes.

Finally, we note that a paper with similar aims to our own appeared during the final stages of preparation
of this manuscript: \cite{Urrestilla:2011gr}. The authors of this work utilize the Abelian Higgs model string template
and compare with data from the ACBAR, QUAD, and ACT experiments, rather than the SPT. The results of \cite{Urrestilla:2011gr}
and ours are thus quite complementary and contain relatively little overlap, despite the similarity in goals that both
works share.

\acknowledgements
We thank  R. Keisler for help replicating the SPT team's
foreground calculations, L. Pogosian for CMBACT support, and
T. Crawford and M. Zaldarriaga for useful discussions.
CD and WH were supported by
 the Kavli Institute for Cosmological Physics (KICP) at the University
 of Chicago through grants NSF PHY-0114422 and NSF PHY-0551142 and an
 endowment from the Kavli Foundation and its founder Fred Kavli.
CD was additionally supported by the Institute for Advanced Study through the NSF grant AST-0807444 and the Raymond and Beverly Sackler Funds.
MW and WH were supported by  U.S.~Dept.\ of Energy contract
 DE-FG02-90ER-40560 and WH additionally by the David and Lucile Packard Foundation.

\bibliography{stringpaper}

\begin{thebibliography}{63}%
\makeatletter
\providecommand \@ifxundefined [1]{%
 \@ifx{#1\undefined}
}%
\providecommand \@ifnum [1]{%
 \ifnum #1\expandafter \@firstoftwo
 \else \expandafter \@secondoftwo
 \fi
}%
\providecommand \@ifx [1]{%
 \ifx #1\expandafter \@firstoftwo
 \else \expandafter \@secondoftwo
 \fi
}%
\providecommand \natexlab [1]{#1}%
\providecommand \enquote  [1]{``#1''}%
\providecommand \bibnamefont  [1]{#1}%
\providecommand \bibfnamefont [1]{#1}%
\providecommand \citenamefont [1]{#1}%
\providecommand \href@noop [0]{\@secondoftwo}%
\providecommand \href [0]{\begingroup \@sanitize@url \@href}%
\providecommand \@href[1]{\@@startlink{#1}\@@href}%
\providecommand \@@href[1]{\endgroup#1\@@endlink}%
\providecommand \@sanitize@url [0]{\catcode `\\12\catcode `\$12\catcode
  `\&12\catcode `\#12\catcode `\^12\catcode `\_12\catcode `\%12\relax}%
\providecommand \@@startlink[1]{}%
\providecommand \@@endlink[0]{}%
\providecommand \url  [0]{\begingroup\@sanitize@url \@url }%
\providecommand \@url [1]{\endgroup\@href {#1}{\urlprefix }}%
\providecommand \urlprefix  [0]{URL }%
\providecommand \Eprint [0]{\href }%
\@ifxundefined \urlstyle {%
  \providecommand \doi  [0]{\begingroup \@sanitize@url \@doi}%
  \providecommand \@doi [1]{\endgroup \@@startlink {\doibase
  #1}doi:\discretionary {}{}{}#1\@@endlink }%
}{%
  \providecommand \doi  [0]{doi:\discretionary{}{}{}\begingroup
  \urlstyle{rm}\Url }%
}%
\providecommand \doibase [0]{http://dx.doi.org/}%
\providecommand \Doi [0]{\begingroup \@sanitize@url \@Doi }%
\providecommand \@Doi  [1]{\endgroup\@@startlink{\doibase#1}\@@Doi}%
\providecommand \@@Doi [1]{#1\@@endlink}%
\providecommand \selectlanguage [0]{\@gobble}%
\providecommand \bibinfo  [0]{\@secondoftwo}%
\providecommand \bibfield  [0]{\@secondoftwo}%
\providecommand \translation [1]{[#1]}%
\providecommand \BibitemOpen [0]{}%
\providecommand \bibitemStop [0]{}%
\providecommand \bibitemNoStop [0]{.\EOS\space}%
\providecommand \EOS [0]{\spacefactor3000\relax}%
\providecommand \BibitemShut  [1]{\csname bibitem#1\endcsname}%
\bibitem [{\citenamefont {Baumann}\ \emph {et~al.}(2009)\citenamefont {Baumann}
  \emph {et~al.}}]{cmbpol}%
  \BibitemOpen
  \bibfield  {author} {\bibinfo {author} {\bibfnamefont {D.}~\bibnamefont
  {Baumann}} \emph {et~al.} (\bibinfo {collaboration} {CMBPol Study Team}),\
  }\Doi {10.1063/1.3160885} {\bibfield  {journal} {\bibinfo  {journal} {AIP
  Conf.Proc.},\ }\textbf {\bibinfo {volume} {1141}},\ \bibinfo {pages} {10}
  (\bibinfo {year} {2009})},\ \Eprint {http://arxiv.org/abs/0811.3919}
  {arXiv:0811.3919 [astro-ph]} \BibitemShut {NoStop}%
\bibitem [{\citenamefont {Komatsu}\ \emph {et~al.}(2009)\citenamefont
  {Komatsu}, \citenamefont {Afshordi}, \citenamefont {Bartolo}, \citenamefont
  {Baumann}, \citenamefont {Bond} \emph {et~al.}}]{nongauss}%
  \BibitemOpen
  \bibfield  {author} {\bibinfo {author} {\bibfnamefont {E.}~\bibnamefont
  {Komatsu}}, \bibinfo {author} {\bibfnamefont {N.}~\bibnamefont {Afshordi}},
  \bibinfo {author} {\bibfnamefont {N.}~\bibnamefont {Bartolo}}, \bibinfo
  {author} {\bibfnamefont {D.}~\bibnamefont {Baumann}}, \bibinfo {author}
  {\bibfnamefont {J.}~\bibnamefont {Bond}},  \emph {et~al.},\ }\href@noop {} {
  (\bibinfo {year} {2009})},\ \Eprint {http://arxiv.org/abs/0902.4759}
  {arXiv:0902.4759 [astro-ph.CO]} \BibitemShut {NoStop}%
\bibitem [{\citenamefont {Polchinski}(2004)}]{Polchinski:2004ia}%
  \BibitemOpen
  \bibfield  {author} {\bibinfo {author} {\bibfnamefont {J.}~\bibnamefont
  {Polchinski}},\ }\href@noop {} {\bibfield  {journal} {\bibinfo  {journal}
  {{2004 Cargese Summer School}},\ \bibinfo {pages} {229}} (\bibinfo {year}
  {2004})},\ \Eprint {http://arxiv.org/abs/hep-th/0412244}
  {arXiv:hep-th/0412244 [hep-th]} \BibitemShut {NoStop}%
\bibitem [{\citenamefont {Myers}\ and\ \citenamefont
  {Wyman}(2009)}]{Myers:1900zz}%
  \BibitemOpen
  \bibfield  {author} {\bibinfo {author} {\bibfnamefont {R.~C.}\ \bibnamefont
  {Myers}}\ and\ \bibinfo {author} {\bibfnamefont {M.}~\bibnamefont {Wyman}},\
  }\href@noop {} {\emph {\bibinfo {title} {{Cosmic superstrings}}}},\ edited
  by\ \bibinfo {editor} {\bibfnamefont {J.}~\bibnamefont {Erdmenger}}\
  (\bibinfo  {publisher} {Wiley-VHC},\ \bibinfo {year} {2009})\BibitemShut
  {NoStop}%
\bibitem [{\citenamefont {Copeland}\ \emph {et~al.}(2011)\citenamefont
  {Copeland}, \citenamefont {Pogosian},\ and\ \citenamefont
  {Vachaspati}}]{Copeland:2011dx}%
  \BibitemOpen
  \bibfield  {author} {\bibinfo {author} {\bibfnamefont {E.~J.}\ \bibnamefont
  {Copeland}}, \bibinfo {author} {\bibfnamefont {L.}~\bibnamefont {Pogosian}},
  \ and\ \bibinfo {author} {\bibfnamefont {T.}~\bibnamefont {Vachaspati}},\
  }\href@noop {} { (\bibinfo {year} {2011})},\ \Eprint
  {http://arxiv.org/abs/1105.0207} {arXiv:1105.0207 [hep-th]} \BibitemShut
  {NoStop}%
\bibitem [{\citenamefont {Cui}\ \emph {et~al.}(2008)\citenamefont {Cui},
  \citenamefont {Martin}, \citenamefont {Morrissey},\ and\ \citenamefont
  {Wells}}]{Cui:2007js}%
  \BibitemOpen
  \bibfield  {author} {\bibinfo {author} {\bibfnamefont {Y.}~\bibnamefont
  {Cui}}, \bibinfo {author} {\bibfnamefont {S.~P.}\ \bibnamefont {Martin}},
  \bibinfo {author} {\bibfnamefont {D.~E.}\ \bibnamefont {Morrissey}}, \ and\
  \bibinfo {author} {\bibfnamefont {J.~D.}\ \bibnamefont {Wells}},\ }\Doi
  {10.1103/PhysRevD.77.043528} {\bibfield  {journal} {\bibinfo  {journal}
  {Phys.Rev.},\ }\textbf {\bibinfo {volume} {D77}},\ \bibinfo {pages} {043528}
  (\bibinfo {year} {2008})},\ \Eprint {http://arxiv.org/abs/0709.0950}
  {arXiv:0709.0950 [hep-ph]} \BibitemShut {NoStop}%
\bibitem [{\citenamefont {Sarangi}\ and\ \citenamefont
  {Tye}(2002)}]{Sarangi:2002yt}%
  \BibitemOpen
  \bibfield  {author} {\bibinfo {author} {\bibfnamefont {S.}~\bibnamefont
  {Sarangi}}\ and\ \bibinfo {author} {\bibfnamefont {S.}~\bibnamefont {Tye}},\
  }\Doi {10.1016/S0370-2693(02)01824-5} {\bibfield  {journal} {\bibinfo
  {journal} {Phys.Lett.},\ }\textbf {\bibinfo {volume} {B536}},\ \bibinfo
  {pages} {185} (\bibinfo {year} {2002})},\ \Eprint
  {http://arxiv.org/abs/hep-th/0204074} {arXiv:hep-th/0204074 [hep-th]}
  \BibitemShut {NoStop}%
\bibitem [{\citenamefont {Albrecht}\ \emph {et~al.}(1997)\citenamefont
  {Albrecht}, \citenamefont {Battye},\ and\ \citenamefont
  {Robinson}}]{Albrecht:1997nt}%
  \BibitemOpen
  \bibfield  {author} {\bibinfo {author} {\bibfnamefont {A.}~\bibnamefont
  {Albrecht}}, \bibinfo {author} {\bibfnamefont {R.~A.}\ \bibnamefont
  {Battye}}, \ and\ \bibinfo {author} {\bibfnamefont {J.}~\bibnamefont
  {Robinson}},\ }\Doi {10.1103/PhysRevLett.79.4736} {\bibfield  {journal}
  {\bibinfo  {journal} {Phys.Rev.Lett.},\ }\textbf {\bibinfo {volume} {79}},\
  \bibinfo {pages} {4736} (\bibinfo {year} {1997})},\ \Eprint
  {http://arxiv.org/abs/astro-ph/9707129} {arXiv:astro-ph/9707129 [astro-ph]}
  \BibitemShut {NoStop}%
\bibitem [{\citenamefont {Pogosian}\ \emph {et~al.}(2003)\citenamefont
  {Pogosian}, \citenamefont {Tye}, \citenamefont {Wasserman},\ and\
  \citenamefont {Wyman}}]{Pogosian:2003mz}%
  \BibitemOpen
  \bibfield  {author} {\bibinfo {author} {\bibfnamefont {L.}~\bibnamefont
  {Pogosian}}, \bibinfo {author} {\bibfnamefont {S.}~\bibnamefont {Tye}},
  \bibinfo {author} {\bibfnamefont {I.}~\bibnamefont {Wasserman}}, \ and\
  \bibinfo {author} {\bibfnamefont {M.}~\bibnamefont {Wyman}},\ }\Doi
  {10.1103/PhysRevD.68.023506, 10.1103/PhysRevD.73.089904} {\bibfield
  {journal} {\bibinfo  {journal} {Phys.Rev.},\ }\textbf {\bibinfo {volume}
  {D68}},\ \bibinfo {pages} {023506} (\bibinfo {year} {2003})},\ \Eprint
  {http://arxiv.org/abs/hep-th/0304188} {arXiv:hep-th/0304188 [hep-th]}
  \BibitemShut {NoStop}%
\bibitem [{\citenamefont {Pogosian}\ \emph {et~al.}(2004)\citenamefont
  {Pogosian}, \citenamefont {Wyman},\ and\ \citenamefont
  {Wasserman}}]{Pogosian:2004ny}%
  \BibitemOpen
  \bibfield  {author} {\bibinfo {author} {\bibfnamefont {L.}~\bibnamefont
  {Pogosian}}, \bibinfo {author} {\bibfnamefont {M.~C.}\ \bibnamefont {Wyman}},
  \ and\ \bibinfo {author} {\bibfnamefont {I.}~\bibnamefont {Wasserman}},\
  }\Doi {10.1088/1475-7516/2004/09/008} {\bibfield  {journal} {\bibinfo
  {journal} {JCAP},\ }\textbf {\bibinfo {volume} {0409}},\ \bibinfo {pages}
  {008} (\bibinfo {year} {2004})},\ \Eprint
  {http://arxiv.org/abs/astro-ph/0403268} {arXiv:astro-ph/0403268 [astro-ph]}
  \BibitemShut {NoStop}%
\bibitem [{\citenamefont {Wyman}\ \emph {et~al.}(2005)\citenamefont {Wyman},
  \citenamefont {Pogosian},\ and\ \citenamefont {Wasserman}}]{Wyman:2005tu}%
  \BibitemOpen
  \bibfield  {author} {\bibinfo {author} {\bibfnamefont {M.}~\bibnamefont
  {Wyman}}, \bibinfo {author} {\bibfnamefont {L.}~\bibnamefont {Pogosian}}, \
  and\ \bibinfo {author} {\bibfnamefont {I.}~\bibnamefont {Wasserman}},\ }\Doi
  {10.1103/PhysRevD.72.023513, 10.1103/PhysRevD.73.089905} {\bibfield
  {journal} {\bibinfo  {journal} {Phys.Rev.},\ }\textbf {\bibinfo {volume}
  {D72}},\ \bibinfo {pages} {023513} (\bibinfo {year} {2005})},\ \Eprint
  {http://arxiv.org/abs/astro-ph/0503364} {arXiv:astro-ph/0503364 [astro-ph]}
  \BibitemShut {NoStop}%
\bibitem [{\citenamefont {Seljak}\ \emph {et~al.}(1997)\citenamefont {Seljak},
  \citenamefont {Pen},\ and\ \citenamefont {Turok}}]{Seljak:1997ii}%
  \BibitemOpen
  \bibfield  {author} {\bibinfo {author} {\bibfnamefont {U.}~\bibnamefont
  {Seljak}}, \bibinfo {author} {\bibfnamefont {U.-L.}\ \bibnamefont {Pen}}, \
  and\ \bibinfo {author} {\bibfnamefont {N.}~\bibnamefont {Turok}},\ }\Doi
  {10.1103/PhysRevLett.79.1615} {\bibfield  {journal} {\bibinfo  {journal}
  {Phys.Rev.Lett.},\ }\textbf {\bibinfo {volume} {79}},\ \bibinfo {pages}
  {1615} (\bibinfo {year} {1997})},\ \Eprint
  {http://arxiv.org/abs/astro-ph/9704231} {arXiv:astro-ph/9704231 [astro-ph]}
  \BibitemShut {NoStop}%
\bibitem [{\citenamefont {Battye}(1998)}]{Battye:1998js}%
  \BibitemOpen
  \bibfield  {author} {\bibinfo {author} {\bibfnamefont {R.}~\bibnamefont
  {Battye}},\ }\href@noop {} { (\bibinfo {year} {1998})},\ \Eprint
  {http://arxiv.org/abs/astro-ph/9806115} {arXiv:astro-ph/9806115 [astro-ph]}
  \BibitemShut {NoStop}%
\bibitem [{\citenamefont {Benabed}\ and\ \citenamefont
  {Bernardeau}(1999)}]{Benabed:1999wn}%
  \BibitemOpen
  \bibfield  {author} {\bibinfo {author} {\bibfnamefont {K.}~\bibnamefont
  {Benabed}}\ and\ \bibinfo {author} {\bibfnamefont {F.}~\bibnamefont
  {Bernardeau}},\ }\href@noop {} {\bibfield  {journal} {\bibinfo  {journal}
  {Phys.Rev.Lett.}} (\bibinfo {year} {1999})},\ \Eprint
  {http://arxiv.org/abs/astro-ph/9906161} {arXiv:astro-ph/9906161 [astro-ph]}
  \BibitemShut {NoStop}%
\bibitem [{\citenamefont {Bevis}\ \emph
  {et~al.}(2007){\natexlab{a}}\citenamefont {Bevis}, \citenamefont {Hindmarsh},
  \citenamefont {Kunz},\ and\ \citenamefont {Urrestilla}}]{Bevis:2007qz}%
  \BibitemOpen
  \bibfield  {author} {\bibinfo {author} {\bibfnamefont {N.}~\bibnamefont
  {Bevis}}, \bibinfo {author} {\bibfnamefont {M.}~\bibnamefont {Hindmarsh}},
  \bibinfo {author} {\bibfnamefont {M.}~\bibnamefont {Kunz}}, \ and\ \bibinfo
  {author} {\bibfnamefont {J.}~\bibnamefont {Urrestilla}},\ }\Doi
  {10.1103/PhysRevD.76.043005} {\bibfield  {journal} {\bibinfo  {journal}
  {Phys. Rev.},\ }\textbf {\bibinfo {volume} {D76}},\ \bibinfo {pages} {043005}
  (\bibinfo {year} {2007}{\natexlab{a}})},\ \Eprint
  {http://arxiv.org/abs/0704.3800} {arXiv:0704.3800 [astro-ph]} \BibitemShut
  {NoStop}%
\bibitem [{\citenamefont {Pogosian}\ and\ \citenamefont
  {Wyman}(2008)}]{Pogosian:2007gi}%
  \BibitemOpen
  \bibfield  {author} {\bibinfo {author} {\bibfnamefont {L.}~\bibnamefont
  {Pogosian}}\ and\ \bibinfo {author} {\bibfnamefont {M.}~\bibnamefont
  {Wyman}},\ }\Doi {10.1103/PhysRevD.77.083509} {\bibfield  {journal} {\bibinfo
   {journal} {Phys.Rev.},\ }\textbf {\bibinfo {volume} {D77}},\ \bibinfo
  {pages} {083509} (\bibinfo {year} {2008})},\ \Eprint
  {http://arxiv.org/abs/0711.0747} {arXiv:0711.0747 [astro-ph]} \BibitemShut
  {NoStop}%
\bibitem [{\citenamefont {Avgoustidis}\ \emph {et~al.}(2011)\citenamefont
  {Avgoustidis}, \citenamefont {Copeland}, \citenamefont {Moss}, \citenamefont
  {Pogosian}, \citenamefont {Pourtsidou} \emph {et~al.}}]{Avgoustidis:2011ax}%
  \BibitemOpen
  \bibfield  {author} {\bibinfo {author} {\bibfnamefont {A.}~\bibnamefont
  {Avgoustidis}}, \bibinfo {author} {\bibfnamefont {E.~J.}\ \bibnamefont
  {Copeland}}, \bibinfo {author} {\bibfnamefont {A.}~\bibnamefont {Moss}},
  \bibinfo {author} {\bibfnamefont {L.}~\bibnamefont {Pogosian}}, \bibinfo
  {author} {\bibfnamefont {A.}~\bibnamefont {Pourtsidou}},  \emph {et~al.},\
  }\href@noop {} { (\bibinfo {year} {2011})},\ \Eprint
  {http://arxiv.org/abs/1105.6198} {arXiv:1105.6198 [astro-ph.CO]} \BibitemShut
  {NoStop}%
\bibitem [{\citenamefont {Damour}\ and\ \citenamefont
  {Vilenkin}(2000)}]{Damour:2000wa}%
  \BibitemOpen
  \bibfield  {author} {\bibinfo {author} {\bibfnamefont {T.}~\bibnamefont
  {Damour}}\ and\ \bibinfo {author} {\bibfnamefont {A.}~\bibnamefont
  {Vilenkin}},\ }\Doi {10.1103/PhysRevLett.85.3761} {\bibfield  {journal}
  {\bibinfo  {journal} {Phys.Rev.Lett.},\ }\textbf {\bibinfo {volume} {85}},\
  \bibinfo {pages} {3761} (\bibinfo {year} {2000})},\ \Eprint
  {http://arxiv.org/abs/gr-qc/0004075} {arXiv:gr-qc/0004075 [gr-qc]}
  \BibitemShut {NoStop}%
\bibitem [{\citenamefont {Olmez}\ \emph {et~al.}(2010)\citenamefont {Olmez},
  \citenamefont {Mandic},\ and\ \citenamefont {Siemens}}]{Olmez:2010bi}%
  \BibitemOpen
  \bibfield  {author} {\bibinfo {author} {\bibfnamefont {S.}~\bibnamefont
  {Olmez}}, \bibinfo {author} {\bibfnamefont {V.}~\bibnamefont {Mandic}}, \
  and\ \bibinfo {author} {\bibfnamefont {X.}~\bibnamefont {Siemens}},\ }\Doi
  {10.1103/PhysRevD.81.104028} {\bibfield  {journal} {\bibinfo  {journal}
  {Phys.Rev.},\ }\textbf {\bibinfo {volume} {D81}},\ \bibinfo {pages} {104028}
  (\bibinfo {year} {2010})},\ \Eprint {http://arxiv.org/abs/1004.0890}
  {arXiv:1004.0890 [astro-ph.CO]} \BibitemShut {NoStop}%
\bibitem [{\citenamefont {Leblond}\ \emph {et~al.}(2009)\citenamefont
  {Leblond}, \citenamefont {Shlaer},\ and\ \citenamefont
  {Siemens}}]{Leblond:2009fq}%
  \BibitemOpen
  \bibfield  {author} {\bibinfo {author} {\bibfnamefont {L.}~\bibnamefont
  {Leblond}}, \bibinfo {author} {\bibfnamefont {B.}~\bibnamefont {Shlaer}}, \
  and\ \bibinfo {author} {\bibfnamefont {X.}~\bibnamefont {Siemens}},\ }\Doi
  {10.1103/PhysRevD.79.123519} {\bibfield  {journal} {\bibinfo  {journal}
  {Phys.Rev.},\ }\textbf {\bibinfo {volume} {D79}},\ \bibinfo {pages} {123519}
  (\bibinfo {year} {2009})},\ \Eprint {http://arxiv.org/abs/0903.4686}
  {arXiv:0903.4686 [astro-ph.CO]} \BibitemShut {NoStop}%
\bibitem [{\citenamefont {Khatri}\ and\ \citenamefont
  {Wandelt}(2008)}]{Khatri:2008zw}%
  \BibitemOpen
  \bibfield  {author} {\bibinfo {author} {\bibfnamefont {R.}~\bibnamefont
  {Khatri}}\ and\ \bibinfo {author} {\bibfnamefont {B.~D.}\ \bibnamefont
  {Wandelt}},\ }\Doi {10.1103/PhysRevLett.100.091302} {\bibfield  {journal}
  {\bibinfo  {journal} {Phys.Rev.Lett.},\ }\textbf {\bibinfo {volume} {100}},\
  \bibinfo {pages} {091302} (\bibinfo {year} {2008})},\ \Eprint
  {http://arxiv.org/abs/0801.4406} {arXiv:0801.4406 [astro-ph]} \BibitemShut
  {NoStop}%
\bibitem [{\citenamefont {Berndsen}\ \emph {et~al.}(2010)\citenamefont
  {Berndsen}, \citenamefont {Pogosian},\ and\ \citenamefont
  {Wyman}}]{Berndsen:2010xc}%
  \BibitemOpen
  \bibfield  {author} {\bibinfo {author} {\bibfnamefont {A.}~\bibnamefont
  {Berndsen}}, \bibinfo {author} {\bibfnamefont {L.}~\bibnamefont {Pogosian}},
  \ and\ \bibinfo {author} {\bibfnamefont {M.}~\bibnamefont {Wyman}},\
  }\href@noop {} {\bibfield  {journal} {\bibinfo  {journal}
  {Mon.Not.Roy.Astron.Soc.},\ }\textbf {\bibinfo {volume} {407}},\ \bibinfo
  {pages} {1116} (\bibinfo {year} {2010})},\ \Eprint
  {http://arxiv.org/abs/1003.2214} {arXiv:1003.2214 [astro-ph.CO]} \BibitemShut
  {NoStop}%
\bibitem [{\citenamefont {Brandenberger}\ \emph {et~al.}(2010)\citenamefont
  {Brandenberger}, \citenamefont {Danos}, \citenamefont {Hernandez},\ and\
  \citenamefont {Holder}}]{Brandenberger:2010hn}%
  \BibitemOpen
  \bibfield  {author} {\bibinfo {author} {\bibfnamefont {R.~H.}\ \bibnamefont
  {Brandenberger}}, \bibinfo {author} {\bibfnamefont {R.~J.}\ \bibnamefont
  {Danos}}, \bibinfo {author} {\bibfnamefont {O.~F.}\ \bibnamefont
  {Hernandez}}, \ and\ \bibinfo {author} {\bibfnamefont {G.~P.}\ \bibnamefont
  {Holder}},\ }\Doi {10.1088/1475-7516/2010/12/028} {\bibfield  {journal}
  {\bibinfo  {journal} {JCAP},\ }\textbf {\bibinfo {volume} {1012}},\ \bibinfo
  {pages} {028} (\bibinfo {year} {2010})},\ \Eprint
  {http://arxiv.org/abs/1006.2514} {arXiv:1006.2514 [astro-ph.CO]} \BibitemShut
  {NoStop}%
\bibitem [{\citenamefont {Hernandez}\ \emph {et~al.}(2011)\citenamefont
  {Hernandez}, \citenamefont {Wang}, \citenamefont {Brandenberger},\ and\
  \citenamefont {Fong}}]{Hernandez:2011ym}%
  \BibitemOpen
  \bibfield  {author} {\bibinfo {author} {\bibfnamefont {O.~F.}\ \bibnamefont
  {Hernandez}}, \bibinfo {author} {\bibfnamefont {Y.}~\bibnamefont {Wang}},
  \bibinfo {author} {\bibfnamefont {R.}~\bibnamefont {Brandenberger}}, \ and\
  \bibinfo {author} {\bibfnamefont {J.}~\bibnamefont {Fong}},\ }\Doi
  {10.1088/1475-7516/2011/08/014} {\bibfield  {journal} {\bibinfo  {journal}
  {JCAP},\ }\textbf {\bibinfo {volume} {1108}},\ \bibinfo {pages} {014}
  (\bibinfo {year} {2011})},\ \Eprint {http://arxiv.org/abs/1104.3337}
  {arXiv:1104.3337 [astro-ph.CO]} \BibitemShut {NoStop}%
\bibitem [{\citenamefont {Mack}\ \emph {et~al.}(2007)\citenamefont {Mack},
  \citenamefont {Wesley},\ and\ \citenamefont {King}}]{Mack:2007ae}%
  \BibitemOpen
  \bibfield  {author} {\bibinfo {author} {\bibfnamefont {K.~J.}\ \bibnamefont
  {Mack}}, \bibinfo {author} {\bibfnamefont {D.~H.}\ \bibnamefont {Wesley}}, \
  and\ \bibinfo {author} {\bibfnamefont {L.~J.}\ \bibnamefont {King}},\ }\Doi
  {10.1103/PhysRevD.76.123515} {\bibfield  {journal} {\bibinfo  {journal}
  {Phys.Rev.},\ }\textbf {\bibinfo {volume} {D76}},\ \bibinfo {pages} {123515}
  (\bibinfo {year} {2007})},\ \Eprint {http://arxiv.org/abs/astro-ph/0702648}
  {arXiv:astro-ph/0702648 [ASTRO-PH]} \BibitemShut {NoStop}%
\bibitem [{\citenamefont {Chernoff}\ and\ \citenamefont
  {Tye}(2007)}]{Chernoff:2007pd}%
  \BibitemOpen
  \bibfield  {author} {\bibinfo {author} {\bibfnamefont {D.~F.}\ \bibnamefont
  {Chernoff}}\ and\ \bibinfo {author} {\bibfnamefont {S.}~\bibnamefont {Tye}},\
  }\href@noop {} { (\bibinfo {year} {2007})},\ \Eprint
  {http://arxiv.org/abs/0709.1139} {arXiv:0709.1139 [astro-ph]} \BibitemShut
  {NoStop}%
\bibitem [{\citenamefont {Keisler}\ \emph {et~al.}(2011)\citenamefont
  {Keisler}, \citenamefont {Reichardt}, \citenamefont {Aird}, \citenamefont
  {Benson}, \citenamefont {Bleem} \emph {et~al.}}]{Keisler:2011aw}%
  \BibitemOpen
  \bibfield  {author} {\bibinfo {author} {\bibfnamefont {R.}~\bibnamefont
  {Keisler}}, \bibinfo {author} {\bibfnamefont {C.}~\bibnamefont {Reichardt}},
  \bibinfo {author} {\bibfnamefont {K.}~\bibnamefont {Aird}}, \bibinfo {author}
  {\bibfnamefont {B.}~\bibnamefont {Benson}}, \bibinfo {author} {\bibfnamefont
  {L.}~\bibnamefont {Bleem}},  \emph {et~al.},\ }\href@noop {} { (\bibinfo
  {year} {2011})},\ \Eprint {http://arxiv.org/abs/1105.3182} {arXiv:1105.3182
  [astro-ph.CO]} \BibitemShut {NoStop}%
\bibitem [{\citenamefont {Pogosian}\ \emph {et~al.}(2009)\citenamefont
  {Pogosian}, \citenamefont {Tye}, \citenamefont {Wasserman},\ and\
  \citenamefont {Wyman}}]{Pogosian:2008am}%
  \BibitemOpen
  \bibfield  {author} {\bibinfo {author} {\bibfnamefont {L.}~\bibnamefont
  {Pogosian}}, \bibinfo {author} {\bibfnamefont {S.~H.~H.}\ \bibnamefont
  {Tye}}, \bibinfo {author} {\bibfnamefont {I.}~\bibnamefont {Wasserman}}, \
  and\ \bibinfo {author} {\bibfnamefont {M.}~\bibnamefont {Wyman}},\ }\Doi
  {10.1088/1475-7516/2009/02/013} {\bibfield  {journal} {\bibinfo  {journal}
  {JCAP},\ }\textbf {\bibinfo {volume} {0902}},\ \bibinfo {pages} {013}
  (\bibinfo {year} {2009})},\ \Eprint {http://arxiv.org/abs/0804.0810}
  {arXiv:0804.0810 [astro-ph]} \BibitemShut {NoStop}%
\bibitem [{\citenamefont {Hlozek}\ \emph {et~al.}(2011)\citenamefont {Hlozek},
  \citenamefont {Dunkley}, \citenamefont {Addison}, \citenamefont {Appel},
  \citenamefont {Bond} \emph {et~al.}}]{Hlozek:2011pc}%
  \BibitemOpen
  \bibfield  {author} {\bibinfo {author} {\bibfnamefont {R.}~\bibnamefont
  {Hlozek}}, \bibinfo {author} {\bibfnamefont {J.}~\bibnamefont {Dunkley}},
  \bibinfo {author} {\bibfnamefont {G.}~\bibnamefont {Addison}}, \bibinfo
  {author} {\bibfnamefont {J.~W.}\ \bibnamefont {Appel}}, \bibinfo {author}
  {\bibfnamefont {J.}~\bibnamefont {Bond}},  \emph {et~al.},\ }\href@noop {} {
  (\bibinfo {year} {2011})},\ \Eprint {http://arxiv.org/abs/1105.4887}
  {arXiv:1105.4887 [astro-ph.CO]} \BibitemShut {NoStop}%
\bibitem [{\citenamefont {Vilenkin}\ and\ \citenamefont
  {Shellard}(2000)}]{VilenkinShellardBook}%
  \BibitemOpen
  \bibfield  {author} {\bibinfo {author} {\bibfnamefont {A.}~\bibnamefont
  {Vilenkin}}\ and\ \bibinfo {author} {\bibfnamefont {E.~P.}\ \bibnamefont
  {Shellard}},\ }\href@noop {} {\emph {\bibinfo {title} {{Cosmic Strings and
  Other Topological Defects}}}}\ (\bibinfo  {publisher} {Cambridge UP},\
  \bibinfo {year} {2000})\BibitemShut {NoStop}%
\bibitem [{\citenamefont {Jeong}\ and\ \citenamefont
  {Smoot}(2005)}]{Jeong:2004ut}%
  \BibitemOpen
  \bibfield  {author} {\bibinfo {author} {\bibfnamefont {E.}~\bibnamefont
  {Jeong}}\ and\ \bibinfo {author} {\bibfnamefont {G.~F.}\ \bibnamefont
  {Smoot}},\ }\Doi {10.1086/428921} {\bibfield  {journal} {\bibinfo  {journal}
  {Astrophys.J.},\ }\textbf {\bibinfo {volume} {624}},\ \bibinfo {pages} {21}
  (\bibinfo {year} {2005})},\ \Eprint {http://arxiv.org/abs/astro-ph/0406432}
  {arXiv:astro-ph/0406432 [astro-ph]} \BibitemShut {NoStop}%
\bibitem [{\citenamefont {Fraisse}\ \emph {et~al.}(2008)\citenamefont
  {Fraisse}, \citenamefont {Ringeval}, \citenamefont {Spergel},\ and\
  \citenamefont {Bouchet}}]{Fraisse:2007nu}%
  \BibitemOpen
  \bibfield  {author} {\bibinfo {author} {\bibfnamefont {A.~A.}\ \bibnamefont
  {Fraisse}}, \bibinfo {author} {\bibfnamefont {C.}~\bibnamefont {Ringeval}},
  \bibinfo {author} {\bibfnamefont {D.~N.}\ \bibnamefont {Spergel}}, \ and\
  \bibinfo {author} {\bibfnamefont {F.~R.}\ \bibnamefont {Bouchet}},\ }\Doi
  {10.1103/PhysRevD.78.043535} {\bibfield  {journal} {\bibinfo  {journal}
  {Phys.Rev.},\ }\textbf {\bibinfo {volume} {D78}},\ \bibinfo {pages} {043535}
  (\bibinfo {year} {2008})},\ \Eprint {http://arxiv.org/abs/0708.1162}
  {arXiv:0708.1162 [astro-ph]} \BibitemShut {NoStop}%
\bibitem [{\citenamefont {Danos}\ and\ \citenamefont
  {Brandenberger}(2010)}]{Danos:2009vv}%
  \BibitemOpen
  \bibfield  {author} {\bibinfo {author} {\bibfnamefont {R.~J.}\ \bibnamefont
  {Danos}}\ and\ \bibinfo {author} {\bibfnamefont {R.~H.}\ \bibnamefont
  {Brandenberger}},\ }\Doi {10.1088/1475-7516/2010/02/033} {\bibfield
  {journal} {\bibinfo  {journal} {JCAP},\ }\textbf {\bibinfo {volume} {1002}},\
  \bibinfo {pages} {033} (\bibinfo {year} {2010})},\ \Eprint
  {http://arxiv.org/abs/0910.5722} {arXiv:0910.5722 [astro-ph.CO]} \BibitemShut
  {NoStop}%
\bibitem [{\citenamefont {Larson}\ \emph {et~al.}(2011)\citenamefont {Larson},
  \citenamefont {Dunkley}, \citenamefont {Hinshaw}, \citenamefont {Komatsu},
  \citenamefont {Nolta} \emph {et~al.}}]{Larson:2010gs}%
  \BibitemOpen
  \bibfield  {author} {\bibinfo {author} {\bibfnamefont {D.}~\bibnamefont
  {Larson}}, \bibinfo {author} {\bibfnamefont {J.}~\bibnamefont {Dunkley}},
  \bibinfo {author} {\bibfnamefont {G.}~\bibnamefont {Hinshaw}}, \bibinfo
  {author} {\bibfnamefont {E.}~\bibnamefont {Komatsu}}, \bibinfo {author}
  {\bibfnamefont {M.}~\bibnamefont {Nolta}},  \emph {et~al.},\ }\Doi
  {10.1088/0067-0049/192/2/16} {\bibfield  {journal} {\bibinfo  {journal}
  {Astrophys.J.Suppl.},\ }\textbf {\bibinfo {volume} {192}},\ \bibinfo {pages}
  {16} (\bibinfo {year} {2011})},\ \Eprint {http://arxiv.org/abs/1001.4635}
  {arXiv:1001.4635 [astro-ph.CO]} \BibitemShut {NoStop}%
\bibitem [{Note1()}]{Note1}%
  \BibitemOpen
  \bibinfo {note} {\protect \url
  {http://www.sfu.ca/~levon/cmbact.html}}\BibitemShut {NoStop}%
\bibitem [{\citenamefont {Pogosian}\ and\ \citenamefont
  {Vachaspati}(1999)}]{Pogosian:1999np}%
  \BibitemOpen
  \bibfield  {author} {\bibinfo {author} {\bibfnamefont {L.}~\bibnamefont
  {Pogosian}}\ and\ \bibinfo {author} {\bibfnamefont {T.}~\bibnamefont
  {Vachaspati}},\ }\Doi {10.1103/PhysRevD.60.083504} {\bibfield  {journal}
  {\bibinfo  {journal} {Phys. Rev.},\ }\textbf {\bibinfo {volume} {D60}},\
  \bibinfo {pages} {083504} (\bibinfo {year} {1999})},\ \Eprint
  {http://arxiv.org/abs/astro-ph/9903361} {arXiv:astro-ph/9903361} \BibitemShut
  {NoStop}%
\bibitem [{\citenamefont {Seljak}\ and\ \citenamefont
  {Zaldarriaga}(1996)}]{Seljak:1996is}%
  \BibitemOpen
  \bibfield  {author} {\bibinfo {author} {\bibfnamefont {U.}~\bibnamefont
  {Seljak}}\ and\ \bibinfo {author} {\bibfnamefont {M.}~\bibnamefont
  {Zaldarriaga}},\ }\Doi {10.1086/177793} {\bibfield  {journal} {\bibinfo
  {journal} {Astrophys. J.},\ }\textbf {\bibinfo {volume} {469}},\ \bibinfo
  {pages} {437} (\bibinfo {year} {1996})},\ \Eprint
  {http://arxiv.org/abs/astro-ph/9603033} {arXiv:astro-ph/9603033} \BibitemShut
  {NoStop}%
\bibitem [{\citenamefont {Martins}\ and\ \citenamefont
  {Shellard}(1996){\natexlab{a}}}]{Martins:1995tg}%
  \BibitemOpen
  \bibfield  {author} {\bibinfo {author} {\bibfnamefont {C.}~\bibnamefont
  {Martins}}\ and\ \bibinfo {author} {\bibfnamefont {E.}~\bibnamefont
  {Shellard}},\ }\Doi {10.1103/PhysRevD.53.575} {\bibfield  {journal} {\bibinfo
   {journal} {Phys.Rev.},\ }\textbf {\bibinfo {volume} {D53}},\ \bibinfo
  {pages} {575} (\bibinfo {year} {1996}{\natexlab{a}})},\ \Eprint
  {http://arxiv.org/abs/hep-ph/9507335} {arXiv:hep-ph/9507335 [hep-ph]}
  \BibitemShut {NoStop}%
\bibitem [{\citenamefont {Martins}\ and\ \citenamefont
  {Shellard}(1996){\natexlab{b}}}]{Martins:1996jp}%
  \BibitemOpen
  \bibfield  {author} {\bibinfo {author} {\bibfnamefont {C.}~\bibnamefont
  {Martins}}\ and\ \bibinfo {author} {\bibfnamefont {E.}~\bibnamefont
  {Shellard}},\ }\Doi {10.1103/PhysRevD.54.2535} {\bibfield  {journal}
  {\bibinfo  {journal} {Phys.Rev.},\ }\textbf {\bibinfo {volume} {D54}},\
  \bibinfo {pages} {2535} (\bibinfo {year} {1996}{\natexlab{b}})},\ \Eprint
  {http://arxiv.org/abs/hep-ph/9602271} {arXiv:hep-ph/9602271 [hep-ph]}
  \BibitemShut {NoStop}%
\bibitem [{\citenamefont {Martins}\ and\ \citenamefont
  {Shellard}(2002)}]{Martins:2000cs}%
  \BibitemOpen
  \bibfield  {author} {\bibinfo {author} {\bibfnamefont {C.}~\bibnamefont
  {Martins}}\ and\ \bibinfo {author} {\bibfnamefont {E.}~\bibnamefont
  {Shellard}},\ }\Doi {10.1103/PhysRevD.65.043514} {\bibfield  {journal}
  {\bibinfo  {journal} {Phys.Rev.},\ }\textbf {\bibinfo {volume} {D65}},\
  \bibinfo {pages} {043514} (\bibinfo {year} {2002})},\ \Eprint
  {http://arxiv.org/abs/hep-ph/0003298} {arXiv:hep-ph/0003298 [hep-ph]}
  \BibitemShut {NoStop}%
\bibitem [{\citenamefont {Allen}\ \emph {et~al.}(1996)\citenamefont {Allen},
  \citenamefont {Caldwell}, \citenamefont {Shellard}, \citenamefont
  {Stebbins},\ and\ \citenamefont {Veeraraghavan}}]{Allen:1996wi}%
  \BibitemOpen
  \bibfield  {author} {\bibinfo {author} {\bibfnamefont {B.}~\bibnamefont
  {Allen}}, \bibinfo {author} {\bibfnamefont {R.~R.}\ \bibnamefont {Caldwell}},
  \bibinfo {author} {\bibfnamefont {E.~P.~S.}\ \bibnamefont {Shellard}},
  \bibinfo {author} {\bibfnamefont {A.}~\bibnamefont {Stebbins}}, \ and\
  \bibinfo {author} {\bibfnamefont {S.}~\bibnamefont {Veeraraghavan}},\ }\Doi
  {10.1103/PhysRevLett.77.3061} {\bibfield  {journal} {\bibinfo  {journal}
  {Phys. Rev. Lett.},\ }\textbf {\bibinfo {volume} {77}},\ \bibinfo {pages}
  {3061} (\bibinfo {year} {1996})},\ \Eprint
  {http://arxiv.org/abs/astro-ph/9609038} {arXiv:astro-ph/9609038} \BibitemShut
  {NoStop}%
\bibitem [{\citenamefont {Allen}\ \emph {et~al.}(1997)\citenamefont {Allen}
  \emph {et~al.}}]{Allen:1997ag}%
  \BibitemOpen
  \bibfield  {author} {\bibinfo {author} {\bibfnamefont {B.}~\bibnamefont
  {Allen}} \emph {et~al.},\ }\Doi {10.1103/PhysRevLett.79.2624} {\bibfield
  {journal} {\bibinfo  {journal} {Phys. Rev. Lett.},\ }\textbf {\bibinfo
  {volume} {79}},\ \bibinfo {pages} {2624} (\bibinfo {year} {1997})},\ \Eprint
  {http://arxiv.org/abs/astro-ph/9704160} {arXiv:astro-ph/9704160} \BibitemShut
  {NoStop}%
\bibitem [{\citenamefont {Landriau}\ and\ \citenamefont
  {Shellard}(2004)}]{Landriau:2003xf}%
  \BibitemOpen
  \bibfield  {author} {\bibinfo {author} {\bibfnamefont {M.}~\bibnamefont
  {Landriau}}\ and\ \bibinfo {author} {\bibfnamefont {E.~P.~S.}\ \bibnamefont
  {Shellard}},\ }\Doi {10.1103/PhysRevD.69.023003} {\bibfield  {journal}
  {\bibinfo  {journal} {Phys. Rev.},\ }\textbf {\bibinfo {volume} {D69}},\
  \bibinfo {pages} {023003} (\bibinfo {year} {2004})},\ \Eprint
  {http://arxiv.org/abs/astro-ph/0302166} {arXiv:astro-ph/0302166} \BibitemShut
  {NoStop}%
\bibitem [{\citenamefont {Battye}\ and\ \citenamefont
  {Moss}(2010)}]{Battye:2010xz}%
  \BibitemOpen
  \bibfield  {author} {\bibinfo {author} {\bibfnamefont {R.}~\bibnamefont
  {Battye}}\ and\ \bibinfo {author} {\bibfnamefont {A.}~\bibnamefont {Moss}},\
  }\Doi {10.1103/PhysRevD.82.023521} {\bibfield  {journal} {\bibinfo  {journal}
  {Phys.Rev.},\ }\textbf {\bibinfo {volume} {D82}},\ \bibinfo {pages} {023521}
  (\bibinfo {year} {2010})},\ \Eprint {http://arxiv.org/abs/1005.0479}
  {arXiv:1005.0479 [astro-ph.CO]} \BibitemShut {NoStop}%
\bibitem [{\citenamefont {Shirokoff}\ \emph {et~al.}(2011)\citenamefont
  {Shirokoff} \emph {et~al.}}]{Shirokoff:2010cs}%
  \BibitemOpen
  \bibfield  {author} {\bibinfo {author} {\bibfnamefont {E.}~\bibnamefont
  {Shirokoff}} \emph {et~al.},\ }\Doi {10.1088/0004-637X/736/1/61} {\bibfield
  {journal} {\bibinfo  {journal} {Astrophys. J.},\ }\textbf {\bibinfo {volume}
  {736}},\ \bibinfo {pages} {61} (\bibinfo {year} {2011})},\ \Eprint
  {http://arxiv.org/abs/1012.4788} {arXiv:1012.4788 [astro-ph.CO]} \BibitemShut
  {NoStop}%
\bibitem [{\citenamefont {Addison}\ \emph {et~al.}(2011)\citenamefont {Addison}
  \emph {et~al.}}]{Addison:2011se}%
  \BibitemOpen
  \bibfield  {author} {\bibinfo {author} {\bibfnamefont {G.~E.}\ \bibnamefont
  {Addison}} \emph {et~al.},\ }\href@noop {} { (\bibinfo {year} {2011})},\
  \Eprint {http://arxiv.org/abs/1108.4614} {arXiv:1108.4614 [astro-ph.CO]}
  \BibitemShut {NoStop}%
\bibitem [{\citenamefont {Ma}\ \emph {et~al.}(2010)\citenamefont {Ma},
  \citenamefont {Zhao},\ and\ \citenamefont {Brown}}]{Ma:2010yb}%
  \BibitemOpen
  \bibfield  {author} {\bibinfo {author} {\bibfnamefont {Y.-Z.}\ \bibnamefont
  {Ma}}, \bibinfo {author} {\bibfnamefont {W.}~\bibnamefont {Zhao}}, \ and\
  \bibinfo {author} {\bibfnamefont {M.~L.}\ \bibnamefont {Brown}},\ }\Doi
  {10.1088/1475-7516/2010/10/007} {\bibfield  {journal} {\bibinfo  {journal}
  {JCAP},\ }\textbf {\bibinfo {volume} {1010}},\ \bibinfo {pages} {007}
  (\bibinfo {year} {2010})},\ \Eprint {http://arxiv.org/abs/1007.2396}
  {arXiv:1007.2396 [astro-ph.CO]} \BibitemShut {NoStop}%
\bibitem [{\citenamefont {Mukherjee}\ \emph {et~al.}(2011)\citenamefont
  {Mukherjee}, \citenamefont {Urrestilla}, \citenamefont {Kunz}, \citenamefont
  {Liddle}, \citenamefont {Bevis} \emph {et~al.}}]{Mukherjee:2010ve}%
  \BibitemOpen
  \bibfield  {author} {\bibinfo {author} {\bibfnamefont {P.}~\bibnamefont
  {Mukherjee}}, \bibinfo {author} {\bibfnamefont {J.}~\bibnamefont
  {Urrestilla}}, \bibinfo {author} {\bibfnamefont {M.}~\bibnamefont {Kunz}},
  \bibinfo {author} {\bibfnamefont {A.~R.}\ \bibnamefont {Liddle}}, \bibinfo
  {author} {\bibfnamefont {N.}~\bibnamefont {Bevis}},  \emph {et~al.},\ }\Doi
  {10.1103/PhysRevD.83.043003} {\bibfield  {journal} {\bibinfo  {journal}
  {Phys.Rev.},\ }\textbf {\bibinfo {volume} {D83}},\ \bibinfo {pages} {043003}
  (\bibinfo {year} {2011})},\ \Eprint {http://arxiv.org/abs/1010.5662}
  {arXiv:1010.5662 [astro-ph.CO]} \BibitemShut {NoStop}%
\bibitem [{\citenamefont {Achucarro}\ and\ \citenamefont
  {Vachaspati}(2000)}]{Achucarro:1999it}%
  \BibitemOpen
  \bibfield  {author} {\bibinfo {author} {\bibfnamefont {A.}~\bibnamefont
  {Achucarro}}\ and\ \bibinfo {author} {\bibfnamefont {T.}~\bibnamefont
  {Vachaspati}},\ }\Doi {10.1016/S0370-1573(99)00103-9} {\bibfield  {journal}
  {\bibinfo  {journal} {Phys. Rept.},\ }\textbf {\bibinfo {volume} {327}},\
  \bibinfo {pages} {347} (\bibinfo {year} {2000})},\ \bibinfo {note}
  {[Phys.Rept.327:427,2000]},\ \Eprint {http://arxiv.org/abs/hep-ph/9904229}
  {arXiv:hep-ph/9904229} \BibitemShut {NoStop}%
\bibitem [{\citenamefont {Urrestilla}\ \emph {et~al.}(2008)\citenamefont
  {Urrestilla}, \citenamefont {Bevis}, \citenamefont {Hindmarsh}, \citenamefont
  {Kunz},\ and\ \citenamefont {Liddle}}]{Urrestilla:2007sf}%
  \BibitemOpen
  \bibfield  {author} {\bibinfo {author} {\bibfnamefont {J.}~\bibnamefont
  {Urrestilla}}, \bibinfo {author} {\bibfnamefont {N.}~\bibnamefont {Bevis}},
  \bibinfo {author} {\bibfnamefont {M.}~\bibnamefont {Hindmarsh}}, \bibinfo
  {author} {\bibfnamefont {M.}~\bibnamefont {Kunz}}, \ and\ \bibinfo {author}
  {\bibfnamefont {A.~R.}\ \bibnamefont {Liddle}},\ }\Doi
  {10.1088/1475-7516/2008/07/010} {\bibfield  {journal} {\bibinfo  {journal}
  {JCAP},\ }\textbf {\bibinfo {volume} {0807}},\ \bibinfo {pages} {010}
  (\bibinfo {year} {2008})},\ \Eprint {http://arxiv.org/abs/0711.1842}
  {arXiv:0711.1842 [astro-ph]} \BibitemShut {NoStop}%
\bibitem [{\citenamefont {Copeland}\ \emph {et~al.}(2004)\citenamefont
  {Copeland}, \citenamefont {Myers},\ and\ \citenamefont
  {Polchinski}}]{Copeland:2003bj}%
  \BibitemOpen
  \bibfield  {author} {\bibinfo {author} {\bibfnamefont {E.~J.}\ \bibnamefont
  {Copeland}}, \bibinfo {author} {\bibfnamefont {R.~C.}\ \bibnamefont {Myers}},
  \ and\ \bibinfo {author} {\bibfnamefont {J.}~\bibnamefont {Polchinski}},\
  }\Doi {10.1088/1126-6708/2004/06/013} {\bibfield  {journal} {\bibinfo
  {journal} {JHEP},\ }\textbf {\bibinfo {volume} {0406}},\ \bibinfo {pages}
  {013} (\bibinfo {year} {2004})},\ \Eprint
  {http://arxiv.org/abs/hep-th/0312067} {arXiv:hep-th/0312067 [hep-th]}
  \BibitemShut {NoStop}%
\bibitem [{\citenamefont {Tye}\ \emph {et~al.}(2005)\citenamefont {Tye},
  \citenamefont {Wasserman},\ and\ \citenamefont {Wyman}}]{Tye:2005fn}%
  \BibitemOpen
  \bibfield  {author} {\bibinfo {author} {\bibfnamefont {S.-H.}\ \bibnamefont
  {Tye}}, \bibinfo {author} {\bibfnamefont {I.}~\bibnamefont {Wasserman}}, \
  and\ \bibinfo {author} {\bibfnamefont {M.}~\bibnamefont {Wyman}},\ }\Doi
  {10.1103/PhysRevD.71.103508, 10.1103/PhysRevD.71.129906} {\bibfield
  {journal} {\bibinfo  {journal} {Phys.Rev.},\ }\textbf {\bibinfo {volume}
  {D71}},\ \bibinfo {pages} {103508} (\bibinfo {year} {2005})},\ \Eprint
  {http://arxiv.org/abs/astro-ph/0503506} {arXiv:astro-ph/0503506 [astro-ph]}
  \BibitemShut {NoStop}%
\bibitem [{\citenamefont {Jackson}\ \emph {et~al.}(2005)\citenamefont
  {Jackson}, \citenamefont {Jones},\ and\ \citenamefont
  {Polchinski}}]{Jackson:2004zg}%
  \BibitemOpen
  \bibfield  {author} {\bibinfo {author} {\bibfnamefont {M.~G.}\ \bibnamefont
  {Jackson}}, \bibinfo {author} {\bibfnamefont {N.~T.}\ \bibnamefont {Jones}},
  \ and\ \bibinfo {author} {\bibfnamefont {J.}~\bibnamefont {Polchinski}},\
  }\Doi {10.1088/1126-6708/2005/10/013} {\bibfield  {journal} {\bibinfo
  {journal} {JHEP},\ }\textbf {\bibinfo {volume} {0510}},\ \bibinfo {pages}
  {013} (\bibinfo {year} {2005})},\ \Eprint
  {http://arxiv.org/abs/hep-th/0405229} {arXiv:hep-th/0405229 [hep-th]}
  \BibitemShut {NoStop}%
\bibitem [{\citenamefont {Avgoustidis}\ and\ \citenamefont
  {Shellard}(2006)}]{Avgoustidis:2005nv}%
  \BibitemOpen
  \bibfield  {author} {\bibinfo {author} {\bibfnamefont {A.}~\bibnamefont
  {Avgoustidis}}\ and\ \bibinfo {author} {\bibfnamefont {E.}~\bibnamefont
  {Shellard}},\ }\Doi {10.1103/PhysRevD.73.041301} {\bibfield  {journal}
  {\bibinfo  {journal} {Phys.Rev.},\ }\textbf {\bibinfo {volume} {D73}},\
  \bibinfo {pages} {041301} (\bibinfo {year} {2006})},\ \Eprint
  {http://arxiv.org/abs/astro-ph/0512582} {arXiv:astro-ph/0512582 [astro-ph]}
  \BibitemShut {NoStop}%
\bibitem [{\citenamefont {Foreman}\ \emph {et~al.}(2011)\citenamefont
  {Foreman}, \citenamefont {Moss},\ and\ \citenamefont
  {Scott}}]{Foreman:2011uj}%
  \BibitemOpen
  \bibfield  {author} {\bibinfo {author} {\bibfnamefont {S.}~\bibnamefont
  {Foreman}}, \bibinfo {author} {\bibfnamefont {A.}~\bibnamefont {Moss}}, \
  and\ \bibinfo {author} {\bibfnamefont {D.}~\bibnamefont {Scott}},\
  }\href@noop {} { (\bibinfo {year} {2011})},\ \Eprint
  {http://arxiv.org/abs/1106.4018} {arXiv:1106.4018 [astro-ph.CO]} \BibitemShut
  {NoStop}%
\bibitem [{\citenamefont {Bevis}\ \emph
  {et~al.}(2007){\natexlab{b}}\citenamefont {Bevis}, \citenamefont {Hindmarsh},
  \citenamefont {Kunz},\ and\ \citenamefont {Urrestilla}}]{Bevis:2006mj}%
  \BibitemOpen
  \bibfield  {author} {\bibinfo {author} {\bibfnamefont {N.}~\bibnamefont
  {Bevis}}, \bibinfo {author} {\bibfnamefont {M.}~\bibnamefont {Hindmarsh}},
  \bibinfo {author} {\bibfnamefont {M.}~\bibnamefont {Kunz}}, \ and\ \bibinfo
  {author} {\bibfnamefont {J.}~\bibnamefont {Urrestilla}},\ }\Doi
  {10.1103/PhysRevD.75.065015} {\bibfield  {journal} {\bibinfo  {journal}
  {Phys.Rev.},\ }\textbf {\bibinfo {volume} {D75}},\ \bibinfo {pages} {065015}
  (\bibinfo {year} {2007}{\natexlab{b}})},\ \Eprint
  {http://arxiv.org/abs/astro-ph/0605018} {arXiv:astro-ph/0605018 [astro-ph]}
  \BibitemShut {NoStop}%
\bibitem [{\citenamefont {Bevis}\ \emph {et~al.}(2010)\citenamefont {Bevis},
  \citenamefont {Hindmarsh}, \citenamefont {Kunz},\ and\ \citenamefont
  {Urrestilla}}]{Bevis:2010gj}%
  \BibitemOpen
  \bibfield  {author} {\bibinfo {author} {\bibfnamefont {N.}~\bibnamefont
  {Bevis}}, \bibinfo {author} {\bibfnamefont {M.}~\bibnamefont {Hindmarsh}},
  \bibinfo {author} {\bibfnamefont {M.}~\bibnamefont {Kunz}}, \ and\ \bibinfo
  {author} {\bibfnamefont {J.}~\bibnamefont {Urrestilla}},\ }\Doi
  {10.1103/PhysRevD.82.065004} {\bibfield  {journal} {\bibinfo  {journal}
  {Phys.Rev.},\ }\textbf {\bibinfo {volume} {D82}},\ \bibinfo {pages} {065004}
  (\bibinfo {year} {2010})},\ \Eprint {http://arxiv.org/abs/1005.2663}
  {arXiv:1005.2663 [astro-ph.CO]} \BibitemShut {NoStop}%
\bibitem [{Note2()}]{Note2}%
  \BibitemOpen
  \bibinfo {note} {N.B., beyond these theoretical concerns, the finite-width
  spectra also have significantly less power at high-$\ell $ as compared with
  zero-width spectra \cite {Battye:2010xz, Bevis:2007gh}. This will lead to
  less significant improvement to the constraints on strings as compared with
  what we find; see a recent and closely related work, \cite
  {Urrestilla:2011gr}}\BibitemShut {NoStop}%
\bibitem [{\citenamefont {Jenet}\ \emph {et~al.}(2006)\citenamefont {Jenet},
  \citenamefont {Hobbs}, \citenamefont {van Straten}, \citenamefont
  {Manchester}, \citenamefont {Bailes} \emph {et~al.}}]{Jenet:2006sv}%
  \BibitemOpen
  \bibfield  {author} {\bibinfo {author} {\bibfnamefont {F.~A.}\ \bibnamefont
  {Jenet}}, \bibinfo {author} {\bibfnamefont {G.}~\bibnamefont {Hobbs}},
  \bibinfo {author} {\bibfnamefont {W.}~\bibnamefont {van Straten}}, \bibinfo
  {author} {\bibfnamefont {R.}~\bibnamefont {Manchester}}, \bibinfo {author}
  {\bibfnamefont {M.}~\bibnamefont {Bailes}},  \emph {et~al.},\ }\Doi
  {10.1086/508702} {\bibfield  {journal} {\bibinfo  {journal} {Astrophys.J.},\
  }\textbf {\bibinfo {volume} {653}},\ \bibinfo {pages} {1571} (\bibinfo {year}
  {2006})},\ \Eprint {http://arxiv.org/abs/astro-ph/0609013}
  {arXiv:astro-ph/0609013 [astro-ph]} \BibitemShut {NoStop}%
\bibitem [{\citenamefont {van Haasteren}\ \emph {et~al.}(2011)\citenamefont
  {van Haasteren} \emph {et~al.}}]{vanHaasteren:2011ni}%
  \BibitemOpen
  \bibfield  {author} {\bibinfo {author} {\bibfnamefont {R.}~\bibnamefont {van
  Haasteren}} \emph {et~al.},\ }\href@noop {} { (\bibinfo {year} {2011})},\
  \Eprint {http://arxiv.org/abs/1103.0576} {arXiv:1103.0576 [astro-ph.CO]}
  \BibitemShut {NoStop}%
\bibitem [{\citenamefont {Polchinski}(2007)}]{Polchinski:2007qc}%
  \BibitemOpen
  \bibfield  {author} {\bibinfo {author} {\bibfnamefont {J.}~\bibnamefont
  {Polchinski}},\ }\href@noop {} {\bibfield  {journal} {\bibinfo  {journal}
  {{Proceedings of the 11th Marcel Grossman Meeting}},\ \bibinfo {pages} {105}}
  (\bibinfo {year} {2007})},\ \Eprint {http://arxiv.org/abs/0707.0888}
  {arXiv:0707.0888 [astro-ph]} \BibitemShut {NoStop}%
\bibitem [{\citenamefont {Urrestilla}\ \emph {et~al.}(2011)\citenamefont
  {Urrestilla}, \citenamefont {Bevis}, \citenamefont {Hindmarsh},\ and\
  \citenamefont {Kunz}}]{Urrestilla:2011gr}%
  \BibitemOpen
  \bibfield  {author} {\bibinfo {author} {\bibfnamefont {J.}~\bibnamefont
  {Urrestilla}}, \bibinfo {author} {\bibfnamefont {N.}~\bibnamefont {Bevis}},
  \bibinfo {author} {\bibfnamefont {M.}~\bibnamefont {Hindmarsh}}, \ and\
  \bibinfo {author} {\bibfnamefont {M.}~\bibnamefont {Kunz}},\ }\href@noop {} {
  (\bibinfo {year} {2011})},\ \Eprint {http://arxiv.org/abs/1108.2730}
  {arXiv:1108.2730 [astro-ph.CO]} \BibitemShut {NoStop}%
\bibitem [{\citenamefont {Bevis}\ \emph {et~al.}(2008)\citenamefont {Bevis},
  \citenamefont {Hindmarsh}, \citenamefont {Kunz},\ and\ \citenamefont
  {Urrestilla}}]{Bevis:2007gh}%
  \BibitemOpen
  \bibfield  {author} {\bibinfo {author} {\bibfnamefont {N.}~\bibnamefont
  {Bevis}}, \bibinfo {author} {\bibfnamefont {M.}~\bibnamefont {Hindmarsh}},
  \bibinfo {author} {\bibfnamefont {M.}~\bibnamefont {Kunz}}, \ and\ \bibinfo
  {author} {\bibfnamefont {J.}~\bibnamefont {Urrestilla}},\ }\Doi
  {10.1103/PhysRevLett.100.021301} {\bibfield  {journal} {\bibinfo  {journal}
  {Phys.Rev.Lett.},\ }\textbf {\bibinfo {volume} {100}},\ \bibinfo {pages}
  {021301} (\bibinfo {year} {2008})},\ \Eprint
  {http://arxiv.org/abs/astro-ph/0702223} {arXiv:astro-ph/0702223 [ASTRO-PH]}
  \BibitemShut {NoStop}%
\end{thebibliography}%
\end{document}